\documentclass{IEEEtran}[journal,comsoc,twoside]

\PassOptionsToPackage{obeyspaces,hyphens}{url} 

\usepackage{float, enumerate, xcolor, algorithm, algorithmic, tabto, balance, amsmath, graphicx, epstopdf, hyperref, multirow, dblfloatfix, textcomp, listings, eurosym, lstautogobble, multirow, tikz, amsfonts, pdflscape, tabularx, amssymb, booktabs, amssymb, soul, subfigure, chngcntr, longtable, tabularx, graphicx, adjustbox, tikz}

\usepackage[utf8]{inputenc}
\usepackage[T1]{fontenc}
\usepackage[numbers, compress, sort]{natbib}
\usepackage[flushleft]{threeparttable}
\usepackage[most]{tcolorbox}

\newcommand{\layerOne}{{\textit{Layer-1 }}}
\newcommand{\layerTwo}{{\textit{Layer-2 }}}

\newcommand{\negativeSpace}{\kern-0.8em}

\newcommand{\ankit}[1]{\textcolor{black}{#1}}

\usepackage{array}
\newcolumntype{L}{>{\centering\arraybackslash}m{.5\columnwidth}|}

\usepackage{pifont}

\graphicspath{ {./images/} }

\begin{document}
\title{
{
A Survey of Layer-Two Blockchain Protocols
}
}
\author{
Ankit~Gangwal*,
Haripriya~Ravali~Gangavalli$\dagger$, and
Apoorva~Thirupathi$\dagger$ 
\IEEEcompsocitemizethanks{
\IEEEcompsocthanksitem \textit{* Corresponding author: Ankit Gangwal (gangwal@iiit.ac.in)}
}
\thanks{$\dagger$ Both authors contributed equally.}
\thanks{All authors are with International Institute of Information Technology Hyderabad, 500032, India.
}
}

\markboth{
}
{Gangwal \MakeLowercase{\textit{et al.}}: A Survey of Layer-Two Blockchain Protocols}
\maketitle

\begin{abstract}
	After the success of the Bitcoin blockchain, came several cryptocurrencies and blockchain solutions in the last decade. Nonetheless, Blockchain-based systems still suffer from low transaction rates and high transaction processing latencies, which hinder blockchains' scalability. An entire class of solutions, called \layerOne scalability solutions, have attempted to incrementally improve such limitations by adding/modifying fundamental blockchain attributes. Recently, a completely different class of works, called \layerTwo protocols, have emerged to tackle the blockchain scalability issues using unconventional approaches. \layerTwo protocols improve transaction processing rates, periods, and fees by minimizing the use of underlying slow and costly blockchains. In fact, the main chain acts just as an instrument for trust establishment and dispute resolution among \layerTwo participants, where only a few transactions are dispatched to the main chain. Thus, \layerTwo blockchain protocols have the potential to transform the domain. However, rapid and discrete developments have resulted in diverse branches of \layerTwo protocols. In this work, we systematically create a broad taxonomy of such protocols and implementations. We discuss each \layerTwo protocol class in detail and also elucidate their respective approaches, salient features, requirements, etc. Moreover, we outline the issues related to these protocols along with a comparative discussion. Our thorough study will help further systematize the knowledge dispersed in the domain and help the readers to better understand the field of \layerTwo protocols.
\end{abstract}


\IEEEpeerreviewmaketitle

\section{Introduction}
	Blockchain is a digital ledger of assets (e.g., financial transactions) that is typically managed by a network of peer-to-peer nodes. \ankit{It provides a transparent and decentralized approach for publicly-verifiable and tamper-evident record keeping.} Its unique properties help in eliminating the control of a centralized authority, provide ubiquity, and facilitate fairness via its underlying consensus protocol. Blockchain is essentially the fundamental building block of Bitcoin~\cite{nakamoto2008bitcoin}, \ankit{which is a decentralized cryptocurrency - or, simply a digital cash system - towards which researchers have been working over multiple decades~\cite{chaum1, chaum2}.} Blockchain helps in establishing an agreement between mutually distrusting entities even in the absence of a trusted third party. After the success of Bitcoin, a multitude of financial and non-financial fields have been dramatically transformed by the idea of utilizing a blockchain-based distributed public ledger.
	\par
	According to a widely accepted belief, called blockchain trilemma~\cite{conti2019blockchain}, blockchains can prioritize only two features among decentralization, security, and scalability. Decentralization reflects the fundamental nature of a blockchain while security is an absolute requirement. Therefore, achieving scalability has always remained a challenge \ankit{for blockchain researchers and developers.} Even after a decade of its birth, Bitcoin still suffers from high transaction latency and fails to handle transaction load when compared to conventional payment systems. One of the key factors behind \ankit{the} limited scalability of blockchain is directly related to its core working principle, i.e., their underlying consensus protocol. As a representative example, a block in the Bitcoin blockchain can fit only a limited number of transactions while the Bitcoin network adapts itself to generate only one block every ten minutes on average. Such calibrations have severely restricted its transaction throughput to roughly ten Transactions Per Second~(TPS)~\cite{croman2016scaling} while regular payment systems, such as VISA and PayPal, handle thousands of TPS.
	\par
\begin{figure}[H]
	\centering
	\includegraphics[trim = 0mm 550mm 310mm 0mm, clip, width=\linewidth]{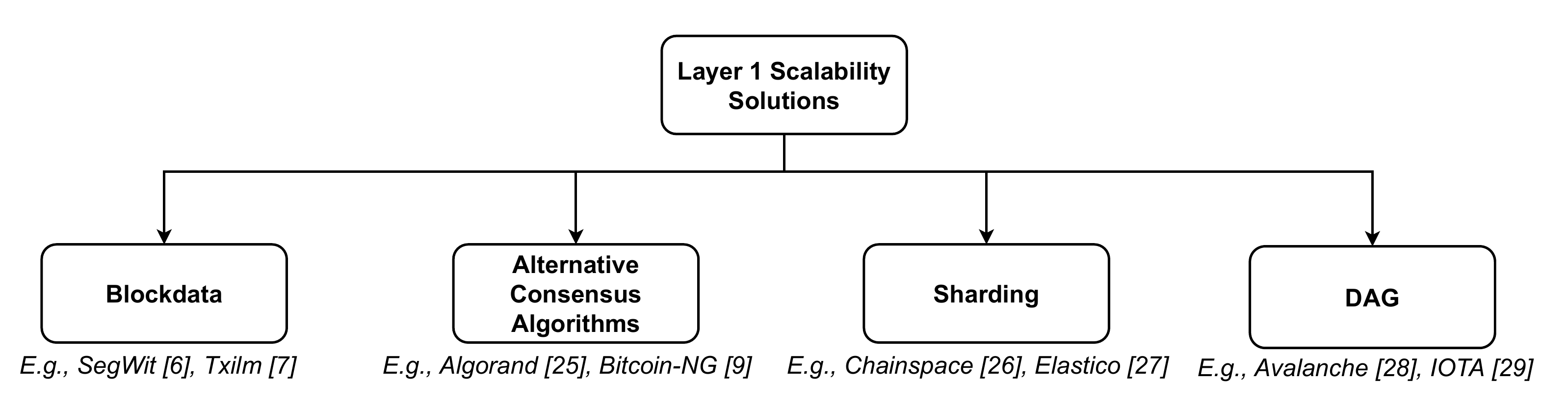}
	\vspace{-1.5em}
	\caption[]{\layerOne scalability solutions.}
	\label{figure:layer_one_solutions}
	\end{figure}

	To tackle the issue of scalability, researchers from both academia and industry have proposed different solutions for scaling blockchains. \ankit{The primary class of such solutions, commonly known as \layerOne solutions, mainly targets and improves the working principles of blockchains by~(i)~modifying block data~\cite{SegWit, Txilm, bip152}; (ii)~ proposing alternative consensus mechanisms~\cite{Bitcoin-ng, pass2017hybrid, kiayias2017ouroboros, Permacoin, Spacemint, Sawtooth0, luu2015scp};
	(iii)~sharding the network~\cite{shardingRoadmap, luu2016secure, gencer2016service, zamani2018rapidchain, kokoris2018omniledger}; or (iv)~using solutions based on Directed Acyclic Graphs~(DAG)~\cite{sompolinsky2016spectre, zhou2019dlattice, cui2019efficient}~(cf. \figurename~\ref{figure:layer_one_solutions}). Since \layerOne solutions involve changing the core design elements of blockchains, these solutions typically lack backward compatibility. As a representative example, modifying the consensus mechanism of a blockchain that is already in-use leads to blockchain forking. Similarly, sharding protocols make significant changes to the overall network layout. Thus, \layerOne solutions come with critical issues that hinder their implementation in practice~\cite{BitcoinCash}.}
\vspace{-1em}
	\phantom{
	SegWit~\cite{SegWit}\\
	Txilm~\cite{Txilm}\\
	Algorand~\cite{Algorand}\\
	Bitcoin-ng~\cite{Bitcoin-ng}\\
	Chainspace~\cite{chainspace}\\
	Elastico~\cite{elastico}\\
	Avalanche~\cite{Avalanche}\\
	IOTA~\cite{IOTA}\\
	}
	\par
\begin{figure}[!htbp]
		\centering
		\includegraphics[trim = 0mm 192mm 130mm 0mm, clip, width=.5\linewidth]{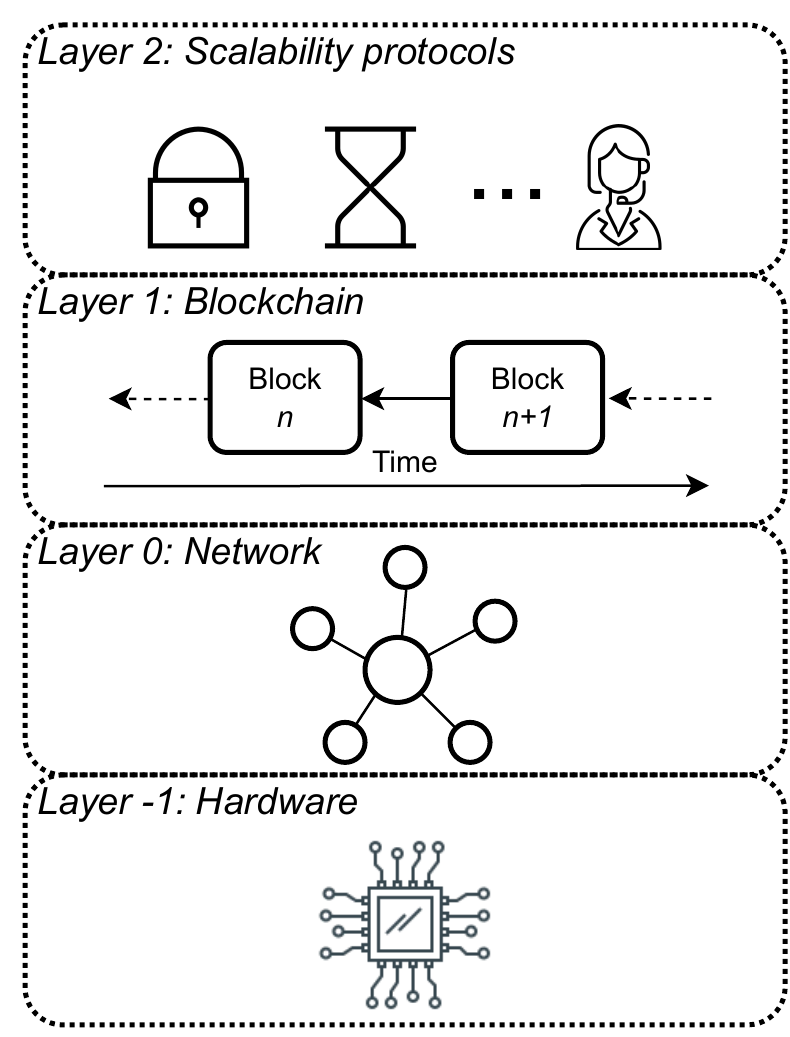}
		\caption[]{Blockchain layered stack, common in blockchain community.}
		\label{figure:layers}
	\end{figure}

	\ankit{The limitations of \layerOne solutions induced an orthogonal research direction for blockchain scalability, generally called \layerTwo blockchain protocols. These protocols aim to scale blockchains without altering the underlying consensus mechanism of the concerned blockchain or modifying the \layerOne trust assumptions.} These solutions are called \layerTwo as they are primarily built over the stack of blockchain layers (cf. \figurename~\ref{figure:layers}), where the	 lowest level (i.e., \textit{layer~-1}) represents the hardware, \textit{layer~0} comprises the network of nodes used for information exchange, the blockchain executes in the \textit{layer~1} of the stack, and \layerTwo scalability solutions sit in \textit{layer~2}. \layerTwo protocols do not broadcast every transaction on the underlying main chain, they instead enable participants to execute off-the-chain transactions over an authenticated communication medium. As a result, transaction load on the main chain is immensely reduced without compromising on backward compatibility. The transactions in \layerTwo protocols are secured with collateral (e.g., in payment channels~\cite{hearn2013micro, bitcoinj, DMC, LightningNetwork}) or with delayed finality (e.g., in commit chains~\cite{khalil2018nocust}).
	\par
	\textit{Motivation:} The unique and promising features of \layerTwo protocols have attracted the attention of researchers in the community. As a result, the research efforts on \layerTwo blockchain scalability protocols are continuing to expand pervasively. Different \layerTwo solutions come with their own set of goals, assumptions, requirements, advantages, etc. There is little clarity - especially for new entrants to the community - to map, understand, and evaluate different \layerTwo solutions. Despite the existence of a rich body of literature on various \layerTwo solutions, a comprehensive study to cover the state of the art detailing their characteristics, limitations, issues, etc. is still missing. Thus, navigating through this research space is not straight forward; especially when the field is growing at a fast pace in different directions. \ankit{We aim to fill such gap in the literature by consolidating and systematizing the information about the state of the art of \layerTwo blockchain protocols.}
	
	\par
	\textit{Contribution:} \ankit{In this paper, we survey various \layerTwo blockchain scalability protocols proposed over the years since the birth of Bitcoin in 2009. Our aim is to present a holistic view of this research field.} To the best of our knowledge, our work is the first such study. 
		We create a broad taxonomy of \layerTwo protocols and implementations and discuss each protocol class in detail. In particular, we explain their respective approaches, key characteristics, advantages, limitations, etc.
		We also discuss the key networking aspects, security concerns, and privacy issues present in the literature.
		Finally, we present a comparative discussion to help readers assess the feasibility of different \layerTwo solutions.

\par
\textit{Organization:} The remainder of this paper is organized as follows. Section~\ref{section:backgroud} covers the key fundamental concepts and related works. We discuss in detail different \layerTwo protocols in Section~\ref{section:layer_two_protocols}. Section~\ref{section:network_issues} describes networking aspects while Section~\ref{section:security_and_privacy_issues} focuses on security and privacy issues. We present a comparative discussion on different \layerTwo scalability solutions in Section~\ref{section:discussion}. Finally, Section~\ref{section:conclusion} concludes the paper.

	\section{Background}
	\label{section:backgroud}
	We introduce the key building blocks of \layerTwo protocols in Section~\ref{subsection:blockchain} and the related works in Section~\ref{subsection:related_work}.
	
	\subsection{Blockchain and HTLC}
	\label{subsection:blockchain}
	Blockchain is an immutable, linked-list style, append-only chain of blocks, where each block stores transactions sent among network entities. Typically, each transaction reflects an exchange of digital assets between network peers. Participants in the network execute a consensus algorithm to achieve a common agreement about the state of the blockchain, which also helps in maintaining its integrity. Today, there are a number of consensus algorithms available. \ankit{Different consensus algorithms follow fundamentally different approaches to achieve distinct goals.} Another key aspect is whether the access to blockchain is open or restricted. In the former case, the blockchain is permissionless while the latter represents a permissioned blockchain. Finally, the scripting language supported by the blockchain defines its expressiveness. As a representative example, \ankit{the} Bitcoin blockchain uses a simple and Turing-incomplete script~\cite{nakamoto2008bitcoin} while Ethereum uses a Turing-complete language to support \ankit{a} more powerful smart contract~\cite{ethereum}. While \layerTwo protocols can be built upon both permissioned and permissionless blockchains, the expressiveness of \ankit{the} underlying blockchain plays an important role in designing  \layerTwo protocols built upon it. Importantly, \layerTwo protocols assume that the underlying blockchain will only include valid transactions to the ledger.
	\par
\begin{figure}[!htbp]
		\centering
		\includegraphics[trim = 0mm 60mm 80mm 0mm, clip, width=.86\linewidth]{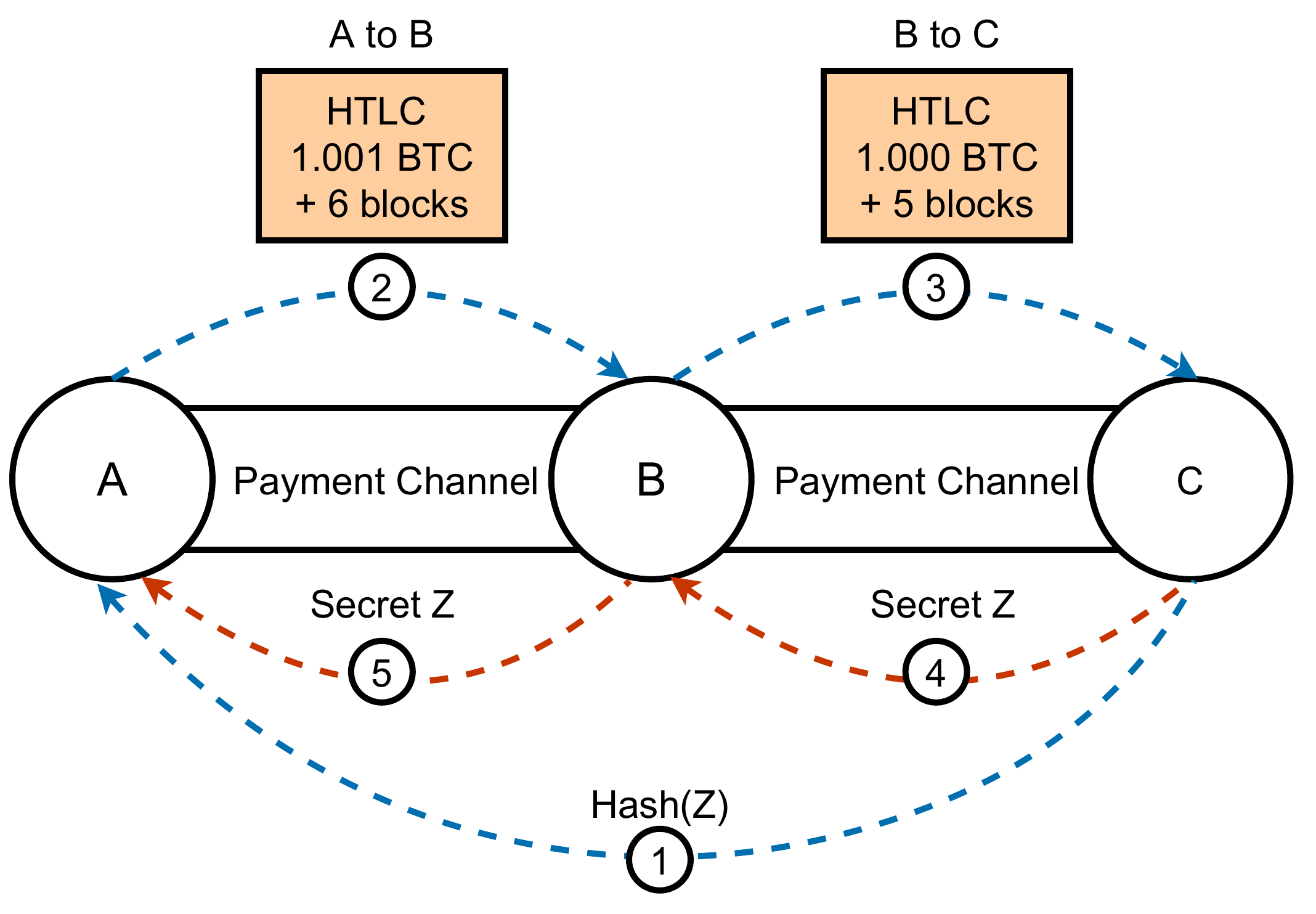}
		\caption[]{A representative example of using HTLC}
		\label{figure:htlc}
	\end{figure}

 The key to a successful P2P transaction system without a central entity relies on a simple and efficient trust-based mechanism. Hash Time Locked Contract~(HTLC)~\cite{HTLC_wiki, DMC, griefing} provides such a solution and acts as the fundamental construction for several \layerTwo protocols. As a representative example, HTLC implementation can be observed in the Lightning network. The core idea of HTLC includes a hash verification and time expiration. 
	\figurename~\ref{figure:htlc} shows an example of using HTLC in payment channels.
	\par
	Consider a payment channel network from A to C through B, where A is the sender and C is the receiver of funds. C generates a secret Z and computes its hash Hash(Z). Then, C communicates Hash(Z) to A. A locks the required funds (including a fee for transfer to intermediaries) in his channel to B and shares Hash(Z) with B in a locking script. B can retrieve the funds only if it can produce Z corresponding to Hash(Z). B further locks funds in his channel with C and passes Hash(Z) to C; B reduces the time to reveal Z and makes some margin on the fee. C can claim these funds only \ankit{if} it can produce Z. Since C knows Z, it can redeem funds in the channel between B and C by revealing Z to B, which can further collect the funds from \ankit{the} channel between B and A. So, intermediaries receive a fee for relaying the transaction. It is worth mentioning that each party can safely participate without \ankit{worrying} about losing their funds as funds frozen in the channel are returned to the original sender when the secret is not produced. Once the secret is revealed, each involved intermediary would work to redeem its payment from the previous channel before the time expiration.

	\subsection{Related works}
	\label{subsection:related_work}
	Several efforts from both academia and industry have been made to address the scalability issues in blockchains. These efforts have targeted different aspects of blockchains starting from finding alternative consensus algorithms, modifying block size, sharding, DAG, \layerTwo protocols, etc. Continuous and rapid \ankit{developments} in this research area have led to many parallel as well as distinct branches of works. Many works have attempted to systematize the knowledge in the rich body of the literature. 
	\par
	\ankit{Several works, such as~\cite{yu2020survey, wang2019sok}, survey scaling solutions that directly engage with the fundamental building blocks of the blockchains. \layerTwo scaling solutions remain out of the scope of such surveys. Only a brief literature on the survey of \layerTwo solutions exists; many of which focus on creating a taxonomy. Authors in~\cite{kim2018survey} classify \layerTwo solutions along a few dimensions. However, their classification omits major protocols such as bisection protocols, commit chains, and TEE~(Trusted Execution Environment)-based solutions. Similarly, the works~\cite{hafid2020scaling, zhou2020solutions} focus on only popular \layerTwo protocols. Authors in~\cite{sok1} discuss various categories of \layerTwo protocols while the work~\cite{sok2} focuses primarily on the network and routing aspects of \layerTwo solutions.}
	\par 
	Our paper aims at furnishing a comprehensive guide for \layerTwo protocols, starting with a much broader taxonomy, a detailed explanation of each protocol, their salient features, and their key concerns. \ankit{To the best of our knowledge, our work covers all the solutions present as of December 2021.}


	\section{Layer-two blockchain protocols}
	\label{section:layer_two_protocols}
\begin{figure*}[!htbp]
	\centering
	\includegraphics[trim = 0mm 202mm 150mm 0mm, clip, width=\linewidth]{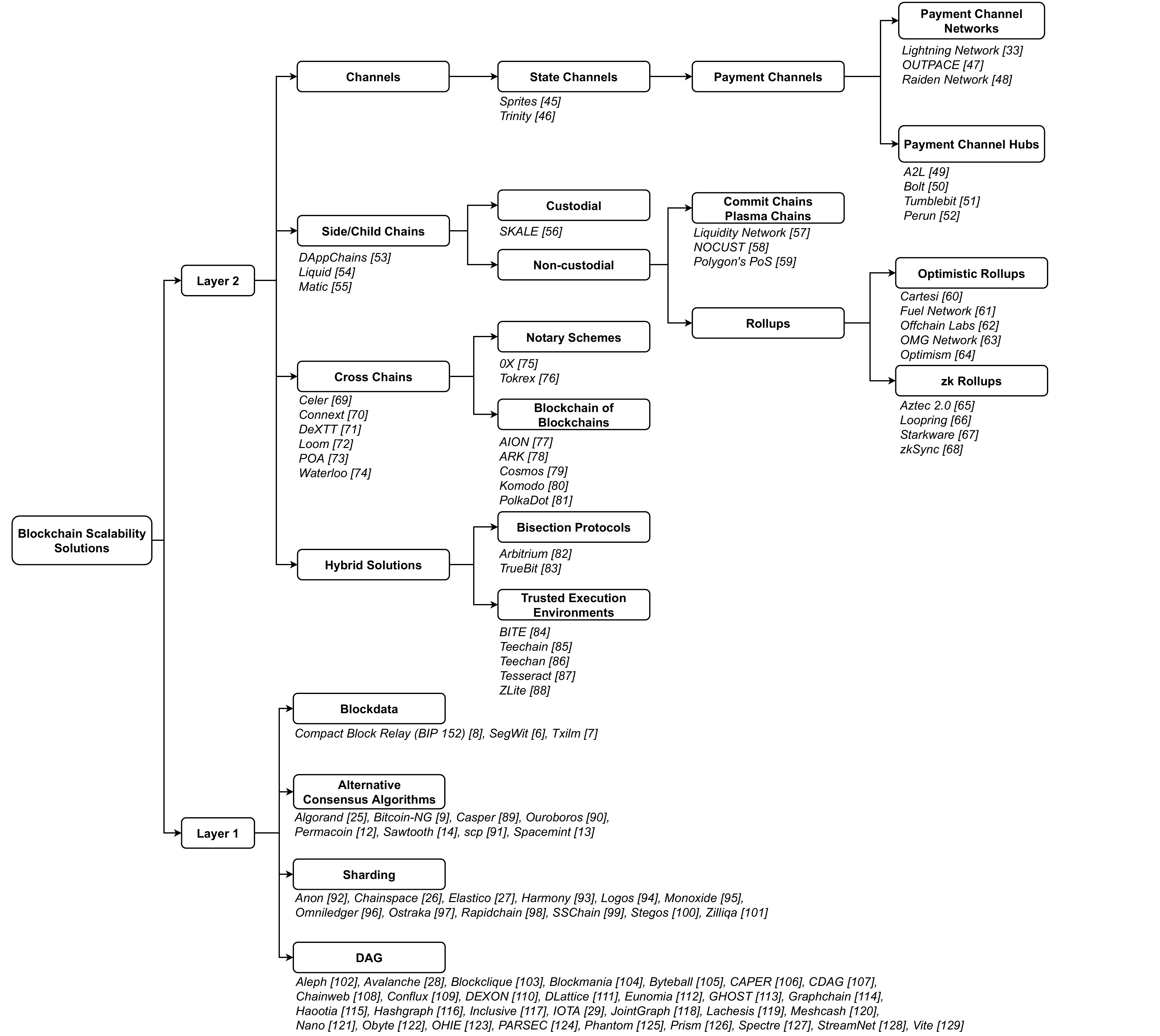}
	\caption[]{Taxonomy of blockchain scalability solutions.}
	\label{figure:layer_one_two_solutions}
	\end{figure*}

	In this section, we elucidate different \layerTwo protocol along with their requirements, working procedures, salient features, etc. In \figurename~\ref{figure:layer_one_two_solutions}, we depict the taxonomy of different blockchain scalability solutions at \layerOne as well as \layerTwo \negativeSpace. Here, a box represents a class/subclass while implementations in a class/subclass are mentioned below respective boxes. These protocols can be broadly categorized into four classes, i.e., channels, side/child chains, cross chains, and hybrid solutions. We now describe each of the categories.
	\phantom{
		Sprites~\cite{Sprites}\\
		Trinity~\cite{Trinity}\\
		~\\%
		Lightning Network~\cite{LightningNetwork}\\
		OUTPACE~\cite{Outpace}\\
		Raiden Network~\cite{RaidenNetwork}\\
		~\\%
		A2L~\cite{A2L}\\
		Bolt~\cite{Bolt}\\
		TumbleBit~\cite{TumbleBit}\\
		Perun~\cite{Perun}\\
		~\\%
		DAppChains~\cite{DAppChains}\\
		Liquid~\cite{Liquid}\\
		Matic~\cite{Matic}\\
		~\\%
		SKALE~\cite{Skale}\\
		~\\%
		Liquidity Network~\cite{LiquidityNetwork}\\
		NOCUST~\cite{NOCUST}\\
		Polygon's PoS~\cite{PolygonsPoS}\\
		~\\%
		Cartesi~\cite{Cartesi}\\
		Fuel Network~\cite{FuelNetwork}\\
		Offchain Labs~\cite{OffchainLabs}\\
		OMG Network~\cite{OMGNetwork}\\
		Optimism~\cite{Optimism}\\
		~\\%
		Aztec 2.0~\cite{Aztec2.0}\\
		Loopring~\cite{Loopring}\\
		Starkware~\cite{Starkware}\\
		zkSync~\cite{zkSync}\\
		~\\%
		Celer~\cite{Celer}\\
		Connext~\cite{Connext}\\
		DeXTT~\cite{DeXTT}\\
		Loom~\cite{Loom}\\
		POA~\cite{POA}\\
		Waterloo~\cite{Waterloo}\\
		~\\%
		0X~\cite{0x}\\
		Tokrex~\cite{Tokrex}\\
		~\\%
		AION~\cite{AION}\\
		ARK~\cite{ARK}\\
		Cosmos~\cite{Cosmos}\\
		Komodo~\cite{Komodo}\\
		PolkaDot~\cite{PolkaDot}\\
		~\\%
		Arbitrum~\cite{Arbitrum}\\
		TrueBit~\cite{TrueBit}\\
		~\\%
		BITE~\cite{BITE}\\
		Teechain~\cite{Teechain}\\
		Teechan~\cite{Teechan}\\
		Tesseract~\cite{Tesseract}\\
		ZLite~\cite{ZLite}\\
		~\\%
		Compact Block Relay (BIP 152)~\cite{bip152}\\
		SegWit~\cite{SegWit}\\
		Txilm~\cite{Txilm}\\
		~\\%
		Algorand~\cite{Algorand}\\
		Bitcoin-ng~\cite{Bitcoin-ng}\\
		Casper~\cite{Casper}\\  
		Ouroboros~\cite{ouroboros}\\
		Permacoin~\cite{Permacoin}\\
		Sawtooth~\cite{Sawtooth0}\\
		scp~\cite{scp}\\
		Spacemint~\cite{Spacemint}\\
		~\\%
		Anon~\cite{anon}\\
		Chainspace~\cite{chainspace}\\
		Elastico~\cite{elastico}\\
		Harmony~\cite{harmony}\\
		Logos~\cite{logos}\\
		Monoxide~\cite{monoxide}\\
		Omniledger~\cite{omniledger}\\
		Ostraka~\cite{ostraka}\\
		Rapidchain~\cite{rapidchain}\\ 
		SSChain~\cite{sschain}\\
		Stegos~\cite{stegos}\\	
		Zilliqa~\cite{zilliqa}\\
		~\\%
		Aleph~\cite{Aleph}\\
		Avalanche~\cite{Avalanche}\\
		Blockclique~\cite{Blockclique}\\
		Blockmania~\cite{Blockmania}\\
		Byteball~\cite{Byteball}\\
		CAPER~\cite{Caper}\\    
		CDAG~\cite{CDAG}\\
		Chainweb~\cite{Chainweb}\\
		Conflux~\cite{Conflux}\\
		DEXON~\cite{DEXON}\\
		DLattice~\cite{DLattice}\\
		Eunomia~\cite{Eunomia}\\
		GHOST~\cite{GHOST}\\
		Graphchain~\cite{Graphchain}\\
		Haootia~\cite{Haootia}\\
		Hashgraph~\cite{Hashgraph}\\
		Inclusive~\cite{Inclusive}\\
		IOTA~\cite{IOTA}\\
		JointGraph~\cite{JointGraph}\\
		Lachesis~\cite{Lachesis}\\
		Meshcash~\cite{Meshcash}\\
		Nano~\cite{Nano}\\
		Obyte~\cite{Obyte}\\
		OHIE~\cite{OHIE}\\
		PARSEC~\cite{PARSEC}\\
		Phantom~\cite{Phantom}\\
		Prism~\cite{Prism}\\
		Spectre~\cite{Spectre}\\
		StreamNet~\cite{StreamNet}\\
		Vite~\cite{Vite}
	}

	\newpage
	\subsection{Channels}
	\label{subsection:channels}
	One of the key \textit{Layer-2} protocols to achieve scalability along with privacy is channels. Channels enable any pair of users to create private mediums for their transactions. The main idea is to enable transactions to happen off the main blockchain and yet maintain the same level of security as an on-chain transaction. For transactions' security\ankit{,} a set of rules are predefined and agreed between the participants. Channels can be mainly classified into two main categories, i.e., state channels and payment channels. State channels (discussed in Section~\ref{subsubsection:state_channels}) are generalized versions while payment channels (discussed in Section~\ref{subsubsection:payment_channels}) are specific to payment-oriented applications. For this reason, we show payment channels in a branch nested from the state channels in~\figurename{~\ref{figure:layer_one_two_solutions}}. Payment channels have been further improvised to form a network and a hub.
	
	\subsubsection{State channels}
	\label{subsubsection:state_channels}
\begin{figure}[H]
		\centering
		\includegraphics[trim = 0mm 60mm 85mm 10mm, clip, width=\linewidth]{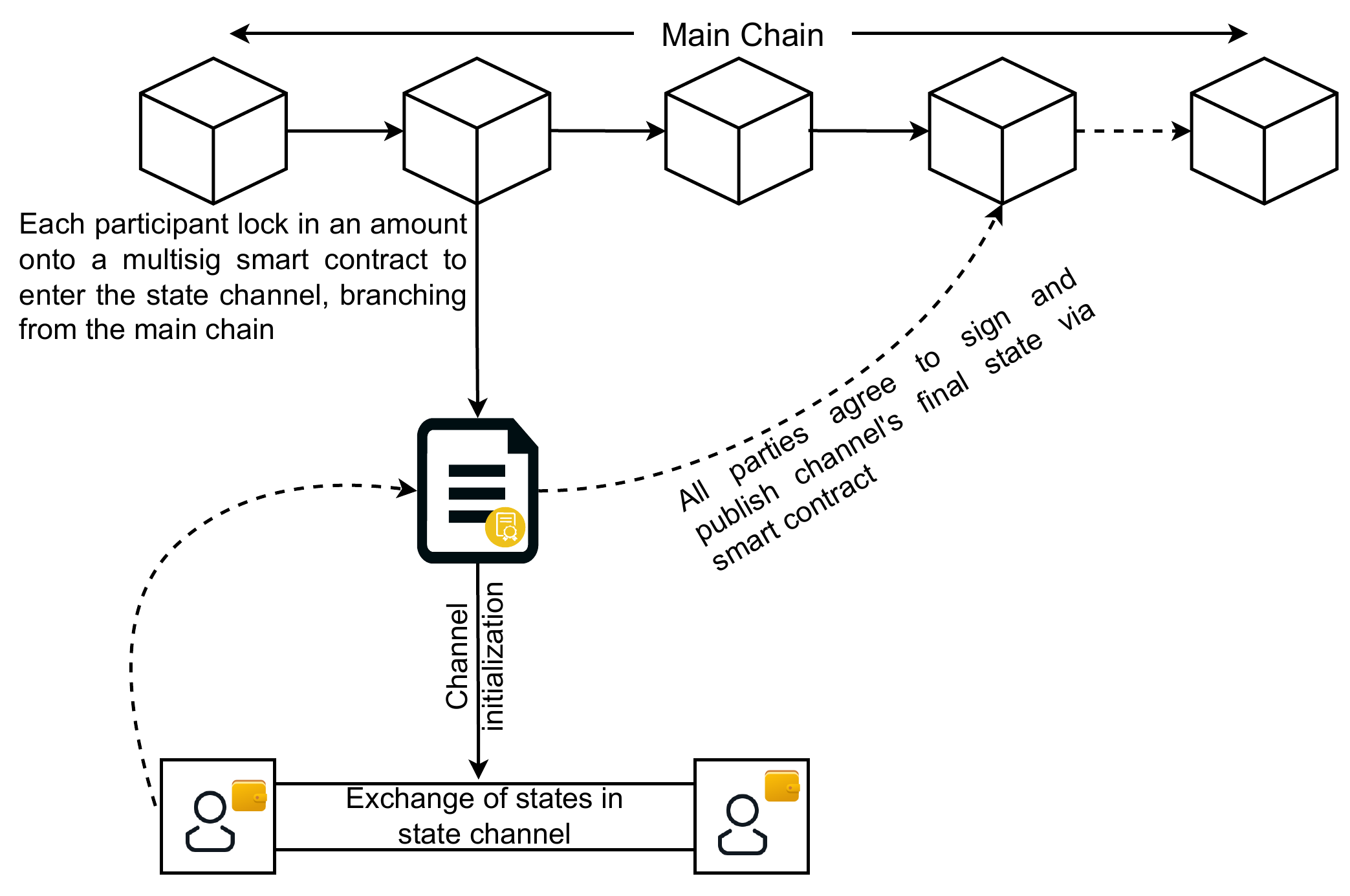}
		\caption[]{Typical lifecycle of a state channel.}
		\label{figure:protocol_stateChannel}
	\end{figure}

	A state channel~\cite{Sprites} is a channel that allows \ankit{the} exchange/transfer of states between two or more participants. These states can represent any arbitrary application (e.g., voting, auctions). Typically, a channel can be built upon threshold signatures - often referred \ankit{to} as \textit{multisig} - and instructions for timelocks~\cite{bip65}, where the participants sign a multisig contract and lock in funds to participate in such a transfer. In practice, state channels are established using \ankit{a} smart contract as shown in \figurename~\ref{figure:protocol_stateChannel}. On these channels, the states are exchanged among all the participants who enter the branched-out channel of states. Once all the transactions \ankit{are} complete, the participants commit the final state of the channel to the main chain via the contract.	
	Such a channel is especially useful when states are exchanged between the participants frequently because off-chain state exchange is far more faster than on-chain exchanges. Thus, state channels greatly improve the slow transaction rate of parent blockchain.\\
	\textit{Lifecycle:} The typical lifecycle of a state channel consists of establishment, execution, and termination phases. To establish the channel, the participants first lock some funds (or assets) using a smart contract on the main chain. The sum of the initially locked funds is the capacity of the established channel. Before proceeding with execution the participants wait for the confirmation of fund locking on the main chain. Moreover, the locked funds can not be used outside the channel. In the execution phase, a transaction happens by an exchange of states between participants. A transaction redistributes the funds from the last agreed states. After exchanging states, the involved participants sign and authorize the new states as valid and true. This new state is then shared with other participants and the order of states is logged. Next, a participant publishes the final state of the channel to the smart contract, which verifies the signatures. Now for the duration of a challenge period, participants other than the publisher are allowed to check if the state is correct. In case of dispute, one can publish the appropriate final state disapproving the previously published state. All the other participants are informed about it, and the challenge period restarts. Typically, the state with \ankit{the} highest version number is considered the latest state. The latest state is executed after the challenge period ends to reflect each participants' state.\\
	\textit{State replacement:} A transaction in state channels is essentially equivalent to replacing the old state with the new state. State replacement should ideally happen once for transaction finality. But to accommodate disagreements among participants about the new proposed state, some state replacement techniques offer a dispute mechanism. Overall, there are four main state replacement techniques:
	\begin{itemize}
		\item \textit{Replace-by-Incentive (RbI):} The sender of a transaction signs and announces a new state. The receiver needs to countersign it to accept the announced state. The motivation of the receiver to accept the state is an incentive; a higher incentive converts to higher chances of state acceptance~\cite{hearn2013micro, spilman, RaidenNetwork}.
		\item \textit{Replace-by-Timelock (RbT):} A state has an associated timelock in terms of either absolute or relative blockchain block height. As the blockchain's block height increases with time, the remaining timelock on a state decreases. Before the expiry of timelock, a state can be replaced with a newer state. Intuitively, a state with the lowest timelock gets included in the blockchain before older states. Post timelock expiry, the transactions represented in a state are written to the blockchain and can not be replaced~\cite{DMC}.
		\item \textit{Replace-by-Revocation (RbR):} There might be a situation where participants want to revoke a state submitted to the blockchain. To do so, all the participants must together propose a new state within a time window defined by the parent blockchain~\cite{LightningNetwork}.
		\item \textit{Replace-by-Version (RbV):} Here, the version of a state is represented by an incrementing counter. A higher version number means a newer state. So, a state with a higher number can replace the older state~\cite{Sprites, watchtower3, Perun, mccorry2019you, counterFactual}.
	\end{itemize}
	\par
	RbI and RbT allow the latest state to be inserted into the blockchain only once. The participants in RbR and RbV can invalidate the submitted state via a dispute process of presenting counter-evidence. The dispute process leads to either a closure dispute or a command dispute. A closure dispute proceeds towards closing the channel and resolving the dispute exclusively on \ankit{the} underlying blockchain. After the dispute is raised, the relevant parties provide evidences within a \ankit{fixed} time duration. At the end of the evidence submission step, any one of the participants processes the evidences to resolve the dispute~\cite{Perun, mccorry2019you}.  
	Instead of closing the channel, command disputes execute a set of commands on the parent-chain to resolve disputes. After command execution, the channel resumes its operation off-chain. The blockchain provides a fixed time duration to collect commands from the participants and executes all the collected commands to find a resolution~\cite{Sprites}. Some solutions~\cite{counterFactual, close2018forcemove} extend dispute process expiry time to support \ankit{the} execution of a large number of commands. However, both the dispute resolution mechanisms assume the relevant participants to be always online. Watching services~\cite{watchtower1, watchtower2, watchtower3, watchtower4} help participants subvert the requirement of staying online by taking the responsibility of observing disagreements.\\
	\textit{Advantages:} The main advantage of using state channels is that all the exchanges happen inside the channel. Unlike main chain transactions where each transaction is broadcast, state \ankit{channels} only publish the final states onto the parent-chain offering more privacy. Instant transaction finality is another advantage, i.e., as soon as all the participants authorize a state update, transaction in that state can be safely considered final. Furthermore, state channels are very economical; especially when state updates are expected to happen frequently between participants. It is so as the cost of updating the states inside the channel is cheaper compared to main chain transaction fees.\\
	\textit{Limitations:} State channels are not suggested for scenarios where the participants are not fixed, i.e., the participants come and leave or whose addresses are not known. Essentially, all the participants must open dedicated channels and be present for state exchange to occur. Furthermore, the dispute process induces an always online assumption. Watching services help here, but such services increase the cost for the participants.
	
	\subsubsection{Payment channels}
	\label{subsubsection:payment_channels}
	Enabling blockchains to support (micro-) payments with near-instant confirmation, fewer on-chain transactions, and reduced fees has been one of the major scalability goals~\cite{SP_dmitrienko2017secure, SP_takahashi2019short, PS_hu2018fast, PS_pass2015micropayments}. Payment channels tailor state channels for payment-specific applications. Initially designed to support one-way payments~\cite{hearn2013micro}, payment channels evolved into bi-directional channels~\cite{DMC, LightningNetwork} to empower each participant to send and receive payments.\\
	\textit{Lifecycle:} Similar to state channels, the lifecycle of a payment channel comprises establishment, execution, and termination/dispute of the payment channel. The payer creates a channel by setting an expiration time, a settlement delay, and a public key to verify claims against the channel. The payee checks if the parameters of the payment channel are suitable for its specific requirements, e.g., destination, settlement delay, channel ID, etc. Importantly, there can be multiple channels between the same pair of participants. Thus, it is important to check the attributes of a channel before initiating a payment. The payer creates a signed claim for the required amount of payment in the channel, which it sends to the payee as the payment for goods or services. It is worth noting that this communication happens ``off-ledger'' over a communication medium suitable for the payer and payee. The payee verifies the claim to ensure that the claim amount is greater than or equal to the total value of the services provided. At this point, the payee can release the goods to the payer because the payment has been assured. The payee is now free to redeem a claim for the authorized amount at any point in time. As the claim values are cumulative, redeeming the largest, i.e., the most recent, claim is sufficient for the payee to get the full amount.
	A channel closure request can lead to two scenarios depending upon whether some funds are still remaining \ankit{in} the channel. If the channel has no funds remaining in it, then the channel can close immediately. Otherwise, the request to close the channel serves as an intimation to the payee to redeem any outstanding claims by the end of settlement delay. The channel expires after the settlement delay has elapsed or the planned expiration time for the channel has arrived. Further transactions can only close the channel, returning any unclaimed funds to the payer. However, an expired channel can last indefinitely on the ledger in its expired state because the ledger can not close it with a closure transaction.\\ 
	\textit{Channel extensions:} Payment channels help channel participants to avoid publishing every transaction on the main chain and wait for subsequent confirmations. Thus, the payments are processed faster and finalized instantly. \figurename{~\ref{figure:protocol_paymentChannel} shows a typical bi-directional channel, where two participants transact with each other in either direction after locking funds during channel setup. Closing the channel reflects their respective final state on the main chain.
	\begin{figure}[!htbp]
		\centering
		\includegraphics[trim = 0mm 155mm 75mm 0mm, clip, width=\linewidth]{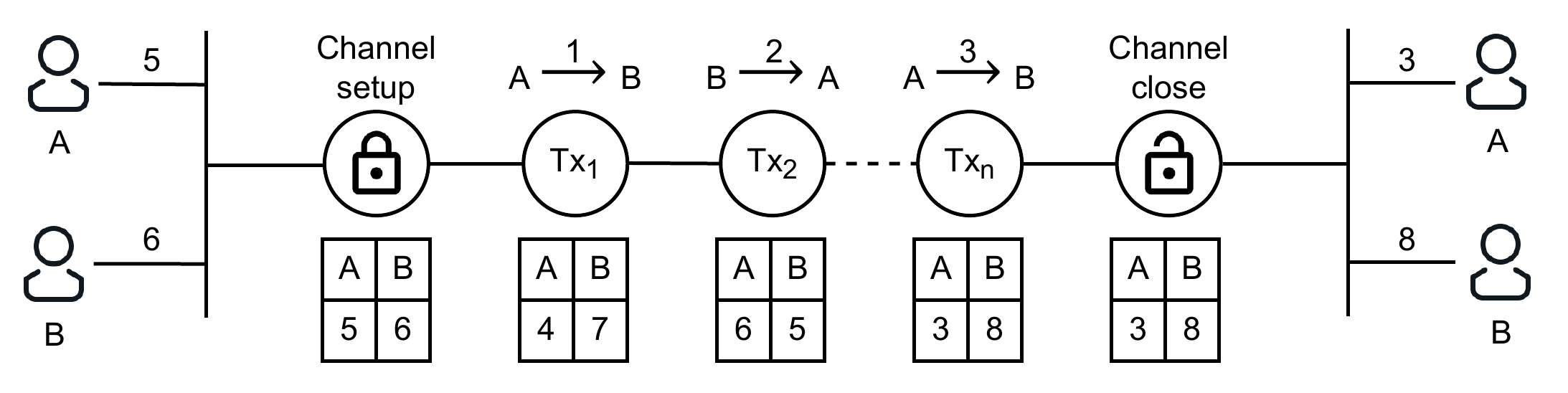}
		\caption[]{A simplified lifecycle of a typical payment channel.}
		\label{figure:protocol_paymentChannel}
	\end{figure}
	\par
	Such an approach is especially useful when the participants frequently transact. However, there are some limitations associated with the payment channels starting with the creation of the channel. Setting up a channel requires locking funds exclusively to the channel. The initial fund locking is also not instant and requires confirmation from the main chain. Moreover, there must be a dedicated channel between the participants. Therefore, such constraints limit the usage of payment channels for micropayments. Nevertheless, several solutions have been proposed to improvise on payment channels, including channel factories, Payment Channel Network~(PCN), payment channel hubs, and virtual channels.
	
	\begin{itemize}
	\item In \textit{channel factories}~\cite{burchert2018scalable, pedrosa2019scalable}, many participants jointly fund a factory. In particular, n participants jointly lock funds in an n-party deposit, which is then used to create payment channels for each pair of depositors. Whenever two participants want to establish a direct channel between them, all depositors update the n-party deposit to re-allocate funds for the new channel. The advantage here is that there is no need to fund and set up separate payment channels for each pair of participants. Nonetheless, opening a factory via n-party fund locking still requires confirmation from the parent chain.
	\item Payment channels in their original form require participants to have a direct link or channel between them. Such requirement limits the potential, and to some extent scalability, of the payment channels. The reason is the practicality (i.e., channel setup delay, locked funds required, etc.) of having a direct channel with many, if not all, participants. \textit{PCNs}~\cite{LightningNetwork} help with the requirement of having a direct channel by creating a network of channels. PCNs have attracted a lot of attention from both academia and industry. The idea behind PCNs is that if A has a channel with B, who has a channel with C. Then, PCN enables A to transact with C via B, i.e., A$\rightarrow$B$\rightarrow$C. B gets an incentive in terms of a small fee for participating in such a transaction.  PCNs primarily utilize conditional payment constructions, such as HTLC~(cf. Section~\ref{subsection:blockchain})~\cite{DMC, LightningNetwork}. The payer conditionally locks the fund of a transaction such that the payee can redeem the funds only if the locking condition is met. Another parameter in this conditional lock is an expiry time, which stimulates faster resolution of the lock by the payee as well as intermediaries. Such conditional transactions must be atomic in nature, meaning that either the transaction should execute completely from the payer to payee or not execute at all. This property helps in providing security for the funds locked by the participating intermediaries~\cite{A2L, malavolta2017concurrency, egger2019atomic}. In HTLC-based solutions, the overall amount of funds locked as collateral along the payment path increases with the length of the payment path. Furthermore, increasing payment length also increases the time for which the funds are reserved. Authors in~\cite{Sprites} use a global PreimageManager smart contract to convert a local channel dispute to a global problem, which helps in reducing collateral locking time. Alternatively, \ankit{the} authors in~\cite{AMHL} introduce a novel cryptographic primitive for channel synchronization that is independent of the parent blockchain's scripting language, and therefore it removes bottlenecks induced by scripts.
	\item \textit{Payment channel hubs}~\cite{Perun, TumbleBit, Bolt} aim to further optimize PCNs by introducing a special node called a hub. A hub acts as the center of a star topology and relays payment to connected nodes. The core idea here is to reduce routing overheads and funds locked by individual nodes in PCNs~\cite{avarikioti2018payment}. Multiple inter-connected hubs in a network can lead to reduced routing length, and consequently, reduced routing cost and collateral cost at each channel. However, the total funds required by a hub to lock can grow significantly with the increasing number of channels and transaction volume. The situation can worsen when the transactions flow majorly in one direction, requiring expensive and slow rebalancing operations~\cite{Revive}.
	\item Channel extensions with intermediaries require intermediaries to actively participate in related transactions. Two-party~\cite{dziembowski2018general} and \ankit{multi-party}~\cite{dziembowski2019multi} \textit{virtual channels} relax such requirements. Virtual channels give an illusion to the payer and payee of having a direct channel between them. Virtual channels are established when all intermediaries between the payer and payee lock funds for a fixed time duration. Setting up a virtual channel between a pair of participants comes at the cost of installing a new virtual channel for each intermediary, where each intermediary must oversee its channels' closure. The main advantage of virtual channels is that channels can be created and closed without blockchain interaction~\cite{dziembowski2019multi}.
	\end{itemize}
	\subsection{Side/Child chains}
	\label{subsection:side_child_chains}
	A side chain~\cite{sidechains, singh2020sidechain} is an independent distributed ledger running in parallel to the main chain. Its primary goals include reducing \ankit{the} load on the main chain by transferring computationally heavy work off the chain. It also \ankit{allows} assets to be transferred across different blockchains. A side chain generally \ankit{utilizes} its own consensus mechanism (e.g., proof-of-authority, proof-of-stake) to process transactions. Side chains use a two-way bridge, called a two-way peg, to communicate and exchange funds with the main chain (cf. \figurename~\ref{figure:protocol_sideChain}). The usability of any side chain depends on its ability to swiftly exchange information with the main chain and quickly process the transactions. Typically, side chains use custom block parameters to process transactions efficiently. In what follows, we explain the pegging mechanism used by side chains.
	\begin{figure}[!htbp]
		\centering
		\includegraphics[trim = 12mm 120mm 90mm 0mm, clip, width=\linewidth]{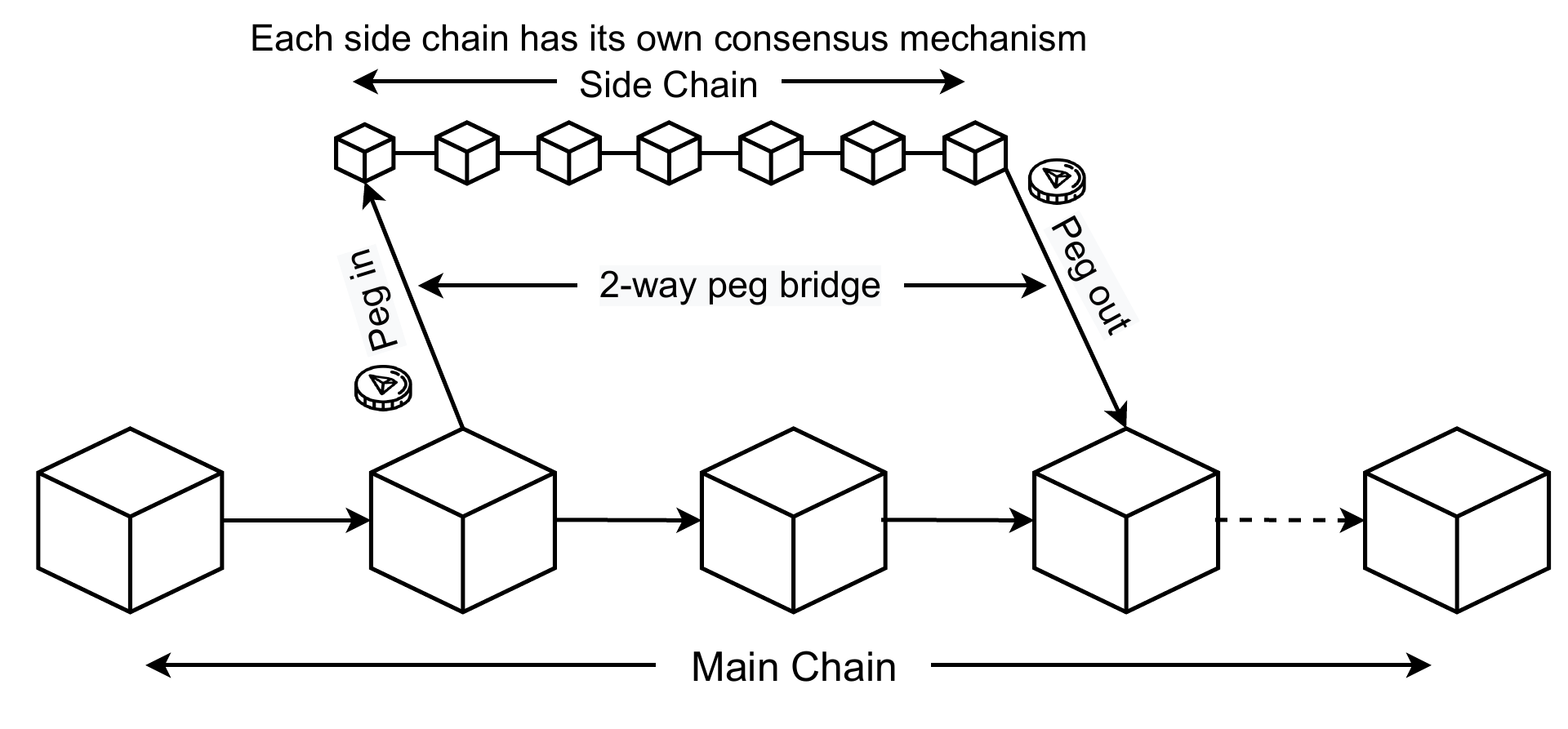}
		\caption[]{A generalized view of side chains.}
		\label{figure:protocol_sideChain}
	\end{figure}~\\
	\textit{Lifecycle:} The two-way peg mechanism allows funds to be transferred between the main chain and side chain at a deterministic exchange rate. Simplified Payment Verification~(SPV) peg is used as the two-way peg. It starts with transferring required funds in the main chain to a special output. Such an output is unlocked by an SPV proof-of-possession inside the side chain. The SPV proof contains a list of block headers showcasing the proof-of-work and a cryptographic proof as evidence (i.e., proof-of-inclusion) that the output was indeed created in one of those blocks. SPV-based pegs enable verifiers to confirm the existence of the special output without downloading the entire main chain. Synchronizing the two chains involves a confirmation period and a contest period. \ankit{The confirmation period corresponds to the time required for the finality of a transaction on the main chain that binds funds to a special output (as mentioned above).} After a finalized SPF proof is created in the main chain, the funds reflect on the side chain in a frozen state. The duration of such freezing is referred to as \ankit{the} contest period, during which new proofs can be published to contest the validity of the locked special output. Contest period helps in preserving \ankit{the} integrity of fund conversion between the main chain and a given side chain. 
	\par
	The funds inside a side chain can move within it without any interaction with the main chain. However, funds remain bonded to the parent chain and can not be transferred further to other chains. Redeeming the funds from a side to the main chain follows the same procedure, i.e., the funds from side chain are locked to a special output, which is then spent using the corresponding SPV proof on the main chain.\\
	\textit{Advantages:} Side chains act as secondary blockchains that provide diverse features and flexibility to their main chains. A side chain has its own independent consensus protocol, and it can control the block parameters. Hence, the transactions on side chains are typically executed faster as compared to main chains. Such processing capabilities also help in reducing the load on the main chain via transaction offloading. Side chains are permanent that can keep running. A new participant can join the same side chain. In contrast, adding participants to the state channel network requires creating a new state channel for each participant. Finally, any compromise or damage remains confined to the side chain only, leaving the main chain unaffected. Such an attribute can be utilized for testing applications before their deployment to the main chain.\\
	\textit{Limitations:} A two-way pegged side chain is slower in execution as a participant needs to wait for a confirmation period as well as a contest period to access the funds on either chain. Another concern is the centralization of mining power on side chains, especially on newer ones. 
	The initial investment required to stabilize the mining process of a side chain and its interoperability with different blockchains constitute the bottleneck of its success. Moreover, the security of funds in a side chain is handled by the side chain. Thus, disputes in side chains are local that can not be resolved in the main chain. 
	\par
	Side chains can be classified into two categories, namely, custodial and non-custodial. Custodial side chains move assets in a chain parallel to the main chain (as explained above) with its own consensus mechanism and security assumptions. On the contrary, the assets and their states are secured via smart contracts on the main chain in non-custodial side chains. 
	Two major classes in non-custodial side chains are Commit (and, Plasma) chains~(discussed in Section~\ref{subsection:commit_chains_plasma}) and Rollups~(discussed in Section~\ref{subsection:rollups}).

	\subsubsection{Commit chains}
	\label{subsection:commit_chains_plasma}
	Channel-based solutions, such as PCN, require participants to open dedicated channels, where funds are locked within the channels and can not be reused elsewhere unless withdrawn from channels; participants must remain online for fund reception, etc. Commit \ankit{chains}~\cite{commitChains, NOCUST} were introduced to address such issues present in channel-based scalability solutions. As shown in \figurename~\ref{figure:protocol_commitChain}, commit chains employ a non-custodial operator that initializes and maintains a commit chain while a smart contract prevents the operator from misbehaving.
	\begin{figure}[H]
		\centering
		\includegraphics[trim = 4mm 75mm 100mm 0mm, clip, width=\linewidth]{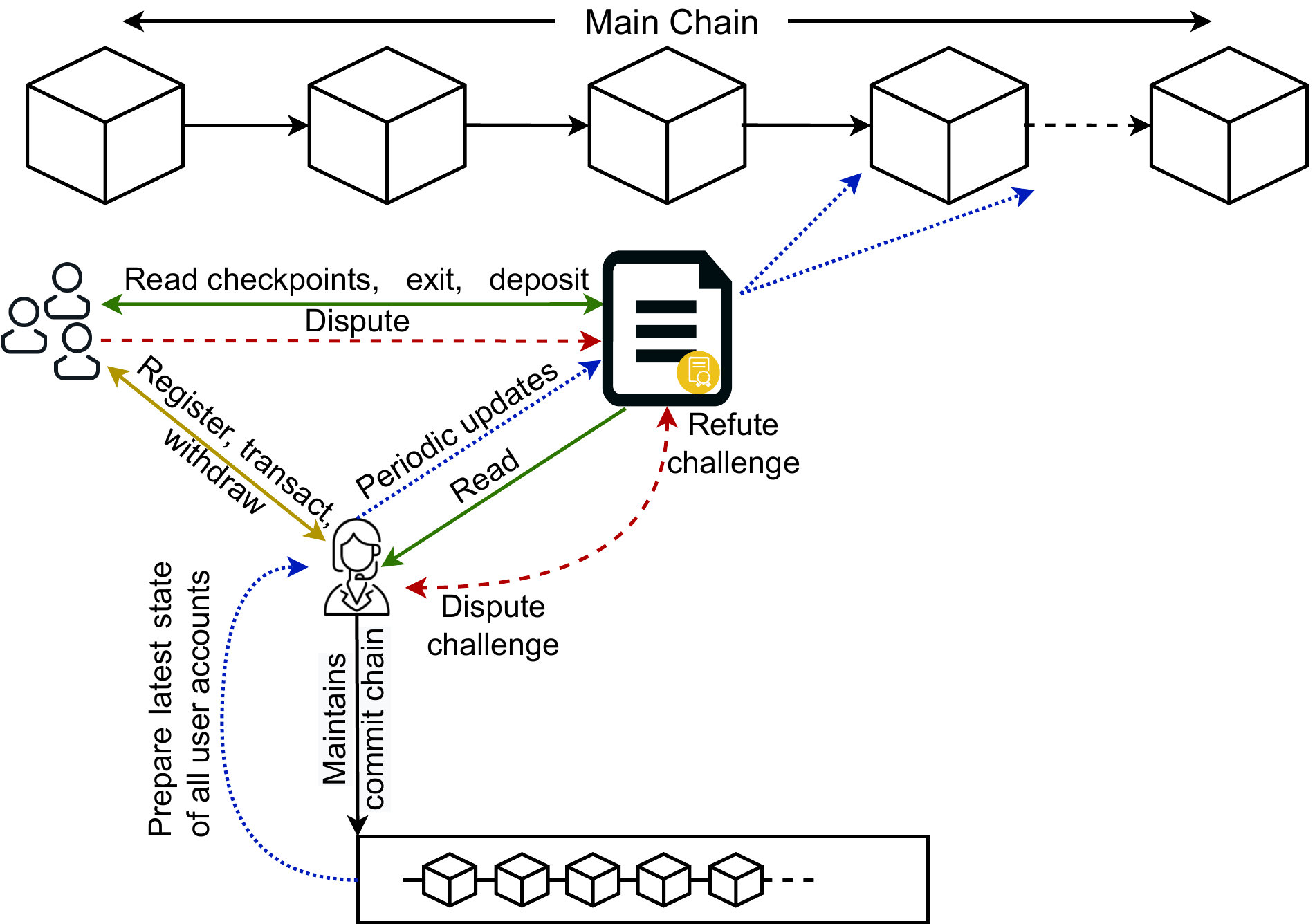}
		\caption[]{An overview of commit chain operations.}
		\label{figure:protocol_commitChain}
	\end{figure}~\\
	\textit{Lifecycle:} Participants willing to join the commit chain first register with the operator to get an account ID on the commit chain. Participants lock funds via the smart contract; the operator reads and updates corresponding account IDs on the commit chain. Recipients do not need to deposit any funds. To transfer funds, the sender authorizes the operator to deduct its account. The operator processes the transactions among the participants off the chain. The operator also periodically commits the latest state of participants' account balances to the main chain through the smart contract using constant-sized checkpoints. The participants observe the checkpoints and challenge the operator via the smart contract in case of a dispute. The smart contract penalizes the operator if found misbehaving, and it also halts the commit chain to recover the balances from the last known stable checkpoint. Finally, a participant can withdraw funds by submitting a withdrawal request to the operator, or it can force exit with the help of the smart contract to close and refund all its account IDs.\\
	\textit{Advantages:} Registering with a commit chain requires no on-chain transaction. Though participants are advised to come online periodically to observe checkpoints, they still receive funds while being offline like any on-chain transactions. Without collateral from an operator, commit chains offer eventual finality. Such an attribute is useful from the operator's perspective. However, an operator may also choose to insure transactions by staking collateral to provide instant finality. Unlike typical side chains, a commit chain is dependent on its parent chain's consensus mechanism, which provides it the same level of security as its parent chain.\\	
	\textit{Limitations:} Although the non-custodial operator is kept in check by the smart contract, it is still the single point of failure. Another issue is that the participants should maintain the commit chain data, which is not published on the parent chain while creating checkpoints, to challenge the operator and exit the commit chain~\cite{NOCUST}.\\
	\textit{Plasma chains:} Another related concept is Plasma~\cite{Plasma} chain. Plasma chains have critical limitations and issues compared to commit chains. Hence, we briefly explain its key concept and concerns. We refer the interested readers to the work~\cite{Plasma} for a detailed description. Commit chain implementations, such as NOCUST~\cite{NOCUST}, are account-based systems. Plasma chain proposes a UTXO-based ledger system running over an account-based blockchain, e.g., Ethereum. Plasma enables multiple blockchains to exist as branches of a tree with the help of a series of smart contracts. Each branch (i.e., blockchain) can have sub-branches (i.e., child chains). Each Plasma chain maintains its own block validation mechanism, which can be independent of its parent chain. However,  all computations in the hierarchy of chains are globally enforced/dependent on one single root chain. Plasma chains suffer from several issues, such as steadily growing data storage costs, high computation requirements, and no native support for instant finality.

	\subsubsection{Rollups}
	\label{subsection:rollups}
	Rollups are non-custodial side chain solutions that aim at reducing the load on the main chain. 
	Rollups employ data compression techniques along with a smart contract for scaling \layerOne chains. The concept is analogous to Plasma chain, except Rollups retain minimal data (in the form of a Merkle root) on-chain about state updates. Such data facilitate on-chain verification and faster withdrawals. The transactions execute off the chain in batches and are bundled together for on-chain verification. In particular, the smart contract maintains the Merkle root (referred to as state root, cf.~\figurename~\ref{figure:protocol_rollup}) on-chain from the current state~(e.g., individual balances) of the Rollup. The same root can also be computed/verified from the data available on-chain. However, the Merkle tree is not stored on-chain to save space. A new state root is computed after a batch of transactions induces balance updates. Anyone can publish the batch by including the transactions in a compressed form, the previous state root, and the newly computed state root. If the current state root in the contract matches the previous state root mentioned in the new batch, the contract updates its state root to the new state root. If a batch requires external inputs (or outputs), the required funds are transferred to the contract before processing (or sent to outputs after processing) the transactions.
	\begin{figure}[H]
	 	\centering
	 	\includegraphics[trim = 0mm 60mm 90mm 2mm, clip, width=\linewidth]{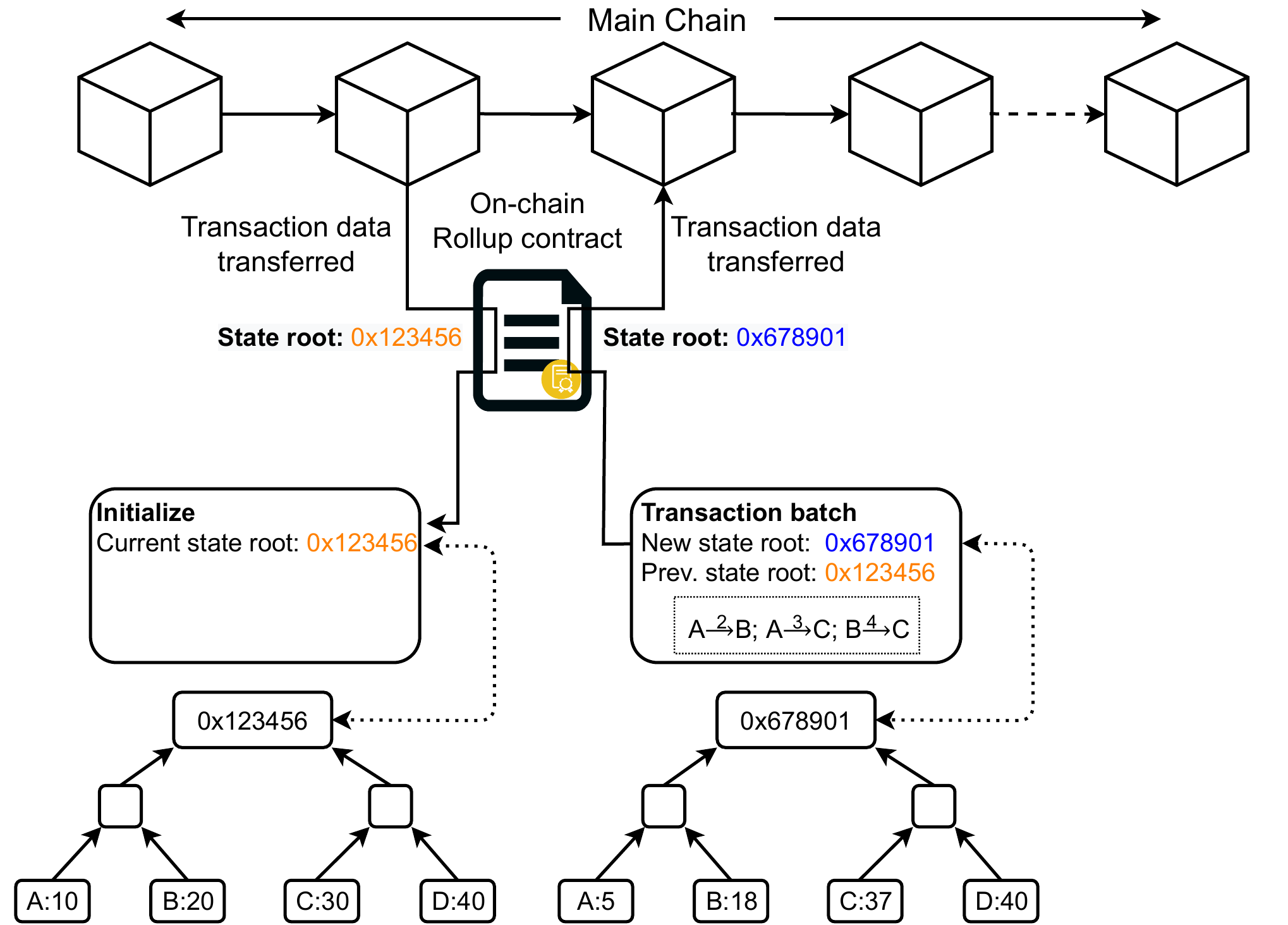}
	 	\caption[]{An abstract presentation of Rollups.}
	 	\label{figure:protocol_rollup}
	\end{figure}
	\par
	Since anyone can publish the batch of transactions, the process of preventing frauds and verifying the new state root leads to two types of rollups, namely Optimistic and Zero-Knowledge~(zk) Rollups.
	\begin{itemize}
		\item \textit{Optimistic Rollups:} Such Rollups take an optimistic approach and assume that transactions are valid unless challenged. Thus, no computation for verification is done by default to significantly improve scalability. However, the contract maintains a history of state root updates along with corresponding batch hashes. Challenging a batch 
		requires publishing a proof of incorrect computations (called fraud proofs) on-chain, which is verified by the contract. Upon verification, the contract reverts the incorrect batch along with all subsequent batches.
		\item \textit{zk Rollups:} Contrary to Optimistic Rollups, zk~Rollups suspect every transaction. Every batch contains a cryptographic proof (called validity proof), which proves that the new state root indeed matches the output of executing the batch of transactions. Such proofs are constructed using zk-SNARK and PLONK protocol~\cite{PLONK}. Computing validity proofs is complex, but their on-chain verification happens quickly.
	\end{itemize}~\\
\textit{Advantages:} Rollups use compression to reduce transaction footprint on-chain that saves space and scales \ankit{the} \layerOne chain. Another key advantage is their ability to bypass the data availability problem~\cite{DataAvailability} with fraud proofs. Optimistic Rollups are suitable for general-purpose computations while zk~Rollups are fit for simple payment scenarios.\\
\textit{Limitations:} The net throughput of Optimistic Rollups is limited. On another side, the cost and complexity of computing validity proofs for zk~Rollups \ankit{are} high.

	\subsection{Cross Chains}
	\label{subsection: cross_chains}
	A huge number of fundamentally different blockchains - in terms of consensus, goals, features, etc. - emerged after the success of the Bitcoin blockchain. Apart from scalability, another major issue with a multitude of blockchains is interoperability. Interoperability not only propels application portability and flexibility further, but it can also assist in improving blockchain scalability by offloading transactions from one blockchain to another~\cite{bcInteroperable}.
	\par
	Cross chains are used to transfer assets between different blockchains. Basically, they facilitate a communication medium between different independent blockchains. Different blockchains have different consensus mechanisms. Transferring assets between a blockchain with weak consensus to a blockchain with strong consensus can lead to potential safety risks. Cross chains help in establishing a mutual trust procedure between users on different blockchains who are interested in swapping assets. In essence, cross chains act as the intermediary platform for inter-blockchain transactions~\cite{XCLAIM, tian2021enabling}. 
	\par
	Celer Network~\cite{Celer} employs a layered architecture with clear abstraction for quick development of generalized state channel as well as side chain suites to support rapid off-chain state transition. Celer network sits between the blockchain and the decentralized applications (``dApps''). Its main component, cStack, is an off-chain technology that supports different blockchains. cStack consists of three main components: cChannel, cRoute, and cOS. cChannel comprises generalized state channels for transactions. cRoute handles routing. cOS is the core component that handles several tasks including resolving dependencies between on-chain and off-chain states, node failures, \ankit{and} unified implementations of on-/off-chain modules. Celer network aims to bring Internet scale to every blockchain while supporting high availability and stable liquidity. Other earlier cross chains are based on HTLC~\cite{DeXTT}, 	relays~(or, side chains)~\cite{POA,Waterloo}, 
	smart contracts~\cite{Connext}. However, there are two state-of-the-art approaches to realize cross chains, i.e., notary scheme~\cite{0x, Tokrex} and blockchain of blockchains~\cite{AION, ARK, Cosmos, Komodo, PolkaDot}.	
	\subsubsection{Notary schemes}
	\label{subsubsection:notary}
	A notary is an entity that actively observes multiple blockchains and listens for transaction events, such as smart contract execution on a chain. It creates a transaction in a chain when a corresponding event happens on another chain. Crypto exchanges, e.g., Binance and Coinbase, are examples of such notary schemes. In practice, exchanges maintain order books to match sellers and buyers. Here, two distrusting parties form an agreement indirectly through the notary. Exchanges that handle and execute trades on behalf of a customer - by holding the customer's private keys - are centralized exchanges while typical decentralized exchanges only offer match-making services.	Decentralized exchanges, such as 0x~\cite{0x}, take a smart contract-based approach (called automated market makers) to constitute a real-time price-adjustment system that replaces the on-chain order books.

	\subsubsection{Blockchain of blockchains}
	\label{subsubsection:bob}
	\ankit{A} blockchain of blockchains, or simply the Internet of blockchains, prioritizes customizability along with interoperability. The general idea here is to build an ecosystem, where independent blockchains can share data and/or tokens with each other via a backbone chain. The backbone chain only facilitates a platform for inter-chain communication programmatically and does not act as the central entity. The customizability perspective emphasizes shortening the blockchain development cycle from years to months~\cite{Cosmos}.
	To summarize, a blockchain of blockchains provides a platform for reusing network, data, incentive, consensus, and layer of contracts to tailor customized, application-specific, interoperable blockchains. Two prominent solutions in this domain are Cosmos~\cite{Cosmos} and Polkadot~\cite{PolkaDot}.
	\par
	Cosmos creates a decentralized network of independent blockchains. These blockchains are called ``zones'', where each zone can have its own constraints on its assets. To enable transactions between zones, Cosmos follows a bridge-hub model. There are multiple hubs present in the network, and each hub can connect multiple zones. Hubs help in reducing the number of connections required to connect different zones. Registering with a hub enables a zone to communicate with all the other zones connected via this hub. Zones communicate with each other only via hubs using Cosmos's Inter-Blockchain Communication~(IBC) protocol. A hub acts as a mutually trusted intermediary that enables zones to share updates regarding their states with other zones. Polkadot introduces globally coherent dynamic data structures called ``parachains'' hosted in parallel while the main chain is called the ``relay'' chain. State transition validation is performed by relay-chain validators. Polkadot defines Cross-Consensus Message Format~(XCM) and Cross-Chain Message Passing~(XCMP) for inter-parachain communication. Polkadot supports connecting hundreds of parachains directly to the relay chain, but only for a short to medium-term. Long-term connections and nested parachains are still under development. A detailed comparison of different cross chain solutions can be found in work~\cite{bcInteroperable}.
	
	\subsection{Hybrid solutions}
	\label{subsection:hybrid_solutions}
	Hybrid solutions help in further improving the scalability of off-chain protocols. These solutions are called hybrid because they change a few fundamental properties of off-chain solutions. We identify two categories of such solutions, where one aims to reduce on-chain dependence of dispute resolution mechanisms while the other uses secure execution mechanisms to eliminate trust requirements among peers. The former category is called bisection protocols (discussed in Section~\ref{subsubsection:bisection}) while the latter (explained in Section~\ref{subsubsection:TEE_solutions}) is implemented using Trusted Execution Environments~(TEE).
	\subsubsection{Bisection protocols}
	\label{subsubsection:bisection}
	Existing dispute resolution mechanisms in off-chain solutions typically execute on-chain. Thus, these solutions are not purely off-chain, at least from the dispute handling perspective. Bisection protocols form a branch of \layerTwo solutions that primarily aim at improving the dispute resolution mechanism. These protocols take part of the computations off-chain, thus helping to reduce the load on the main chain. Generally, bisection protocols involve two steps. First, a user presents minimal evidence to a verifier to testify the validity of its transaction. Next, when users contradict each other, a verifier inspects evidence from contradicting users to determine the correct state. Truebit~\cite{TrueBit} and Arbitrum~\cite{Arbitrum} employ such dispute resolution \ankit{mechanisms}. 
	\par	
	Truebit was introduced as a blockchain enhancement to improve the computational efficiency of smart contracts at a reduced cost. It ports computations from the main chain to \ankit{the} off-chain. Truebit depends on \textit{judges}, who have limited computational power. These judges are mutually trusted by all the participants. Given a computational problem, the user who solves it, called the \textit{solver}, publishes the solution as well as the sub-problems used to reach the solution. A challenge period is set aside during which other users can challenge the published solution - such a user is called a \textit{challenger}. If a challenge is raised, the judges solve the sub-problems recursively using binary search; this is to reduce the problem size by half in each iteration. The solution provided by the judges is mutually agreed to be the correct solution by all participants. The judges compare their correct solution with the solutions provided by the solver and the challenger to identify and penalize the cheating participant.
	\par On another side, Arbitrum uses a Virtual Machine~(VM) to implement a smart contract. A user can create a VM and designate other users as VM managers. An honest manager enforces the VM to follow VM's coded functionality. Any change in the state of the VM has to be approved by all the managers. There might be scenarios where managers do not agree with each other about the state of the VM. Such disagreements should be raised within a challenge period after a new state has been committed. In case of a conflict among the managers, the verifiers/miners invoke a bisection protocol to reduce the conflict down to single-instruction execution. Now, managers present their outcome for that single-instruction execution. The verifiers can verify the presented outcomes efficiently to identify and punish the cheating participant. 
	
	\subsubsection{TEE-based solutions}
	\label{subsubsection:TEE_solutions}
	A trusted execution environment, e.g., Intel SGX~\cite{intelSGX, costan2016intel}, is typically an isolated and safeguarded area inside a CPU, where the integrity and confidentiality of loaded data are protected. TEE-based solutions for blockchain scalability utilize integrity protection offered by TEEs to eliminate the on-chain collateral used for establishing trust among participants. In fact, TEE is used as a mutually trusted entity in such solutions because they offer a higher level of security for application execution.
	\par Teechan\cite{Teechan} uses TEEs to enable two mutually distrusting nodes to transact with each other. Here, a channel is set up between the two nodes by exchanging secrets via their TEEs. As long as the channel is open, the nodes can exchange funds with each other in a peer-to-peer manner using TEE-supported operations; even without involving the parent Bitcoin blockchain. The TEEs bear the responsibility to maintain and update the channel's state securely throughout the channel's lifetime. Upon channel termination, the TEEs create a Bitcoin transaction to be added to the parent chain. During the entire lifetime of such a channel, only two transactions \ankit{are} reflected on-chain; one for channel establishment using a 2-of-2 multisig Bitcoin address and the other for channel closure. To summarize, Teechan reduces the load on the parent chain and increases transaction throughput among distrusting nodes. 
	\par
	 Teechain~\cite{Teechain} is another such solution that executes off-chain transactions asynchronously with the main chain. Teechain employs TEE-protected \textit{treasuries} to preserve the correct channel state. Teechain forms a chain of committee that holds replicated states of treasuries to handle treasury failures.
	\par
	Some other prominent solutions leveraging features of TEEs are Tesseract~\cite{Tesseract}, BITE~\cite{BITE}, and ZLiTE~\cite{ZLite}. Tesseract is a TEE-based cryptocurrency exchange, BITE focuses on the privacy of Bitcoin lightweight clients, and ZLiTE improves the privacy of Zcash lightweight clients by involving TEE-equipped servers. Nevertheless, all TEE-based solutions rely upon the integrity of the TEEs while TEEs have their own vulnerabilities and concerns~\cite{SGXsurvey1, SGXsurvey2}.

	\section{Network issues}
	\label{section:network_issues}
	\layerTwo protocols are built on top of \layerOne blockchains. 
	Enabling participants to transact with each other off the chain may require an overhead communication network to help, for instance, finding a payment path between participants that do not have a direct connection. Routing,  re-balancing, stability, and privacy are the key concerns related to such overhead networks. We take the example of channel-based solutions to briefly discuss these issues. We refer the interested readers to works~\cite{sok1, sok2} for an in-depth analysis of network management issues and a comparison of routing algorithms.\\
	\subsubsection{Routing}
	For a transaction to occur between distant participants in a network, a payment path must be established from the payer to the payee involving intermediate nodes. Deciding such a payment route involves various factors, i.e., inter-node channel capacity, length of the route, cost-effectiveness, and availability of the nodes participating in the transaction. Several works on routing algorithms have attempted to find the most efficient path for the transactions. These routing algorithms can be broadly classified into two categories, namely, global routing~(e.g., ~\cite{spiderRouting} and Di Stasi et al.~\cite{diStasi}) and local routing~(e.g., SilentWhispers~\cite{silentwhispers} and SpeedyMurmurs~\cite{speedyMurmur}). The source of the payment in global routing makes use of the global view of the network to find the optimal path. The performance of such source routing depends on the accuracy of the available global view. By pre-computing paths, these routing algorithms attempt to minimize the overall communication overheads and latencies. However, \ankit{their scalability is limited by} the cost of maintaining an accurate global view and path computations on the source. On another side, local routing algorithms utilize the local information and take a greedy approach. Local routing algorithms are inherently scalable but are less efficient due to local optimizations.
	\begin{table*}[b]
		\caption{A summary of major attacks on \layerTwo solutions.}
		\label{table:attack_descriptions}
		\begin{adjustbox}{angle=00}
			\resizebox{\textwidth}{!}{
				\begin{tabular}{|c|m{5cm}|m{3cm}|c|m{6cm}|m{6cm}|}
					\hline
					\textbf{Attack} & \multicolumn{1}{c|}{\textbf{Impact}} & \multicolumn{1}{c|}{\textbf{Affected parties}} & \textbf{\begin{tabular}[c]{@{}c@{}}No. of adversary\\ nodes needed\end{tabular}} & \multicolumn{1}{c|}{\textbf{Setup and launch}} & \multicolumn{1}{c|}{\textbf{Key prevention/mitigation approaches}} \\ \hline
					
					Wormhole~\cite{AMHL} & Intermediaries' funds are temporarily frozen; incurs \ankit{a} loss of useful time; transaction processing reward is stolen. & All nodes along the path between adversary nodes. & Two & Adversary sets up an additional round of communication to share and bypass the HTLC secret. & AMHL \ankit{offers} interoperable, secure, and privacy-preserving cryptographic construction that works in both script-based and scriptless. \\ \hline
					
					\begin{tabular}[c]{@{}c@{}}Flood\\ and loot~\cite{floodLoot}\end{tabular} & Attacker claims disputed transaction; exploits replace by fee policy. & All nodes that agree to open a channel with attacker's source node. & Two & Adversary establishes several channels through victims and sends a multitude of payments. While settling payments attacker's source node forces victims to create several blockchain transactions all at once. & Reduce the maximum number of unresolved HTLCs; allow more time (blocks) to claim HTLCs on \ankit{the} blockchain; use anchor commitment output; use non-replaceable HTLC transactions. \\ \hline
					
					Griefing~\cite{griefing} & Network stalling; capacity exhaustion; may incur channel force-closing fee; eliminating competitors from \ankit{the} network. & All nodes participating in the payment path towards \ankit{the} attacker. & One & Adversary refuses to resolve payments off-chain, locking victims' funds for  entire duration of contract. & Limit the number of incoming channels; faster HTLC resolution; constant payment locktime; incentivizing/punishing nodes; griefing penalty. \\ \hline
					
					\begin{tabular}[c]{@{}c@{}}Time dilation~\cite{timeDilation}\\ (eclipse)\end{tabular} & Victim isolated from the network; feeds blocks to the victim at a slower rate; funds can be stolen. & Primarily, trust-minimized Bitcoin light clients with limited connections. & Multiple & Adversary deploys hundreds of Sybil nodes, opens a payment channel with the victim, then eclipses/time-dilates the victim. & Increase adoption of BIP 157~\cite{bip157}; anonymize peer-to-peer protocols; \ankit{engage} watchtower. \\ \hline
					
					\begin{tabular}[c]{@{}c@{}}Balance\\ lockdown~\cite{perez2020lockdown}\end{tabular} & Adversary gets a dominant position; blocks the victim's ability to act as an intermediary. & Middle nodes in multi-hop payments. Typically, it targets a single node that relays many payments. & One & Adversary aims to disrupt \ankit{the} availability of a victim, opens a channel with the victim, sends self-destined payments that go and come back via the victim. & Increase AER to reduce attack profitability; reduce the maximum length of a route; minimize/forbid loops in a payment route. \\ \hline
					
					\begin{tabular}[c]{@{}c@{}}Balance\\ discovery~\cite{herrera2019difficulty, dam2020improvements}\end{tabular} & Balance between a pair of victim nodes disclosed; nodes' privacy compromised. & The pair of nodes targeted by attacker. & One & Adversary opens a channel with one of the two victim nodes. It tries to disclose the balance between victims by routing invalid payments. The value of payments typically follows a binary search pattern. & Adhere to protocol \ankit{specifications} and prevent payments with values higher than maximum allowed limits; clients should resist closing a channel upon receiving malformed payments. \\ \hline
					
					Congestion~\cite{mizrahi2021congestion} & No monetary gains; stalls or \ankit{paralyzes} network; DoS affects competitors' gains. & All nodes participating in the transaction between attacker's source and destination node. & One & Adversary overloads channels with unresolved requests to block high liquidity channels and disconnect/isolate individual/pair of nodes. & Avoid paths with loops; enforce fast HTLC resolution; reduce route length; limit the number of maximum concurrent payments. \\ \hline
				\end{tabular}
			}
		\end{adjustbox}
	\end{table*}
	\subsubsection{Re-balancing}
	Exhaustion of capacity in payment channels is a recurring problem. A naive and inefficient solution is to close the channel and refund it, which results in at least two transactions. Protocols like REVIVE~\cite{Revive} allow nodes to safely rebalance their skewed channel with the help of funds available in their other existing channels. An untrusted third party, called a leader, is elected to execute the channel rebalancing process. After receiving nodes' requests and preferences, the leader coordinates with the nodes to freeze relevant channels. The frozen channels are now rebalanced using linear programming and commitments signed by all the nodes. After the process, the nodes lose funds from their one or more payment channels to gain equal funds on the others.
	\subsubsection{Stability and privacy}
	Initial \layerTwo solutions, particularly channel-based, require participants to remain online to monitor transactions and handle disputes. To address such a limitation, participants can engage watchtowers to detect discrepancies on their behalf while they are offline. Watchtowers get a fee for their services. On the contrary, watchtowers are penalized from their collateral for failing to report disputes~\cite{cerberus}. A key concern with watchtowers is that a watchtower may be bribed to cheat its customers for mutual gains. If the bribe is higher than the collateral, the watchtower may act dishonestly. On another side, routing payments over a network of nodes may lead to privacy issues, such as disclosing information about the payer and payee. Malovolta et al.~\cite{malavolta2017concurrency} attempt to formally define the privacy requirements over PCNs and implement Fulger and Rayo using Multi-Hop HTLC smart contract to handle transaction concurrency. However, a routing protocol that learns the capacity of payment channels over a period of time can bypass their privacy notions.
	
	\section{Security and privacy issues}
	\label{section:security_and_privacy_issues}
	In this section, we discuss the security and privacy attacks on different \layerTwo protocols. \tablename~\ref{table:attack_descriptions} presents a summary of all such attacks present in the literature. We clarify their setup requirements, impact, affected parties, and key prevention/mitigation techniques.

	\subsection{Wormhole attack}
	\label{subsection:wormhole_attack}
	Wormhole attack~\cite{AMHL} steals rewards of PCN intermediaries. A payment is typically relayed through multiple intermediaries in PCN since a direct channel may not exist between payer and payee. In such a scenario, the intermediaries are aware of their immediate neighbors and may not be aware of other nodes on the path, sometimes even about the payer and payee. An attacker with just two malicious nodes on the path from the payer to payee can launch wormhole attack. The name wormhole denotes how the funds are rerouted through a wormhole channel between the malicious nodes of the attacker. 
	\par In PCN, the payment is locked in a cryptographic lock, i.e.,  HTLC~(cf. Section~\ref{section:backgroud}), which is routed to the payee. Each intermediate node that relays the payment retains a small fee, which is the difference between the amount it will receive from the channel with \ankit{the} previous node and the amount it will pay to the channel with \ankit{the} next node. Once the payee reveals the key to the lock, it is routed back to the payer. While the key travels back to the payer, it sequentially unlocks funds in the channels on the path. However, a malicious node can bypass the key to another malicious \ankit{nodes}, skipping all the nodes/channels between these two malicious node. As a result, the benign nodes assume that the transaction has failed\ankit{, and they} return to their original state while the malicious \ankit{nodes} steal the reward from the benign nodes.
	\begin{figure}[H]
		\centering
		\includegraphics[trim = 25mm 155mm 10mm 10mm, clip, width=\linewidth]{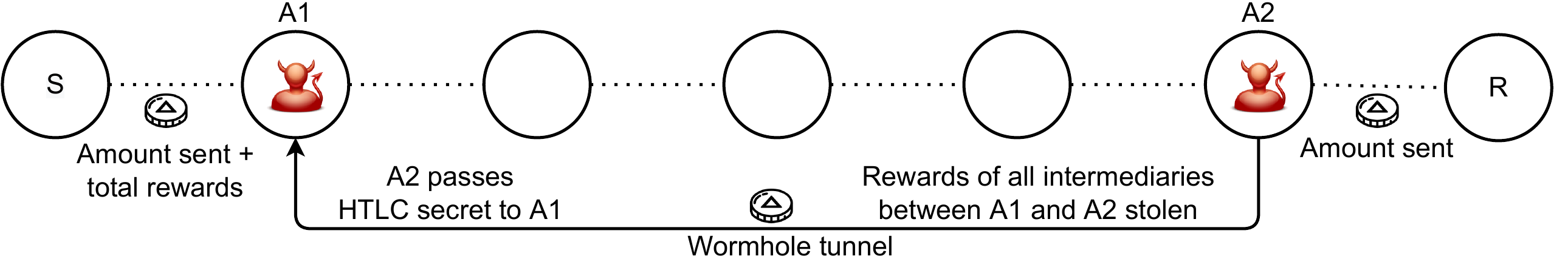}
		\caption[]{Wormhole attack stealing rewards of intermediaries.}
		\label{figure:attack_wormhole}
	\end{figure}
	\figurename{~\ref{figure:attack_wormhole}} depicts a scenario for wormhole attack.  For higher impact, the attack requires the malicious nodes to be closer to the sender (S) and receiver (R). To make a payment from S to R, S locks the original payment amount along with \ankit{the} total reward, i.e., \ankit{the} sum of fees for all intermediaries. Out of all the nodes which agreed to participate in the payment, two nodes A1 and A2 are malicious nodes. When R reveals \ankit{the} HTLC secret to A2, it pays the original payment amount to R. Now A2 bypasses the HTLC secret to A1. A1 uses the HTLC secret to receive the sum of \ankit{the} original payment and the total reward. Thus, A1 and A2 stole the rewards by colluding and bypassing the intermediate nodes, which will return to their original state assuming that the transaction has failed.\\
	\textit{Countermeasures:} The main reasons behind wormhole attack are: (1)~the same HTLC secret is used to unlock funds from each channel on the payment path, and (2)~each channel can be unlocked independently. Anonymous Multi-Hop Lock~(AMHL)~\cite{AMHL} communicates path-specific secret information to nodes and makes locks interdependent to ensure that the funds are unlocked only in a hierarchical manner. 
	
	\subsection{Flood and loot}
	\label{subsection:flood_loot}
	Flood and loot~\cite{floodLoot} is a systemic attack on the Lightning network. The attacker controls two nodes in the network, where one node acts as the source of the payment while the other node is the destination of the payment. Nodes that participate in forming channels between source and target nodes are affected by flood and loot attack. The key idea of the attack is to trigger closing of multiple channels simultaneously. In such a scenario, the victim nodes can only resort to the parent blockchain to claim their funds locked into corresponding open HTLCs. Thus, victim nodes benignly overload the blockchain, which opens a window of opportunity for the attacker to steal the funds. In particular, the attack leverages the replace-by-fee policy~\cite{replaceByFee} employed in \ankit{the} Bitcoin blockchain.
	\par The attacker's source node opens multiple channels through the victim nodes to the attacker's target node. The attacker initiates multiple HTLC payments from the source to target node, \ankit{accepts} these payments at the target node, but refuses to resolve them at the source node. As the timeout for HTLCs approach, victims are compelled to close the channels with the source node and claim open HTLCs on the blockchain.
	\begin{figure}[!htbp]
		\centering
		\subfigure[Establishing channels]{
			\centering
			\includegraphics[trim = 290mm 357mm 130mm 12mm, clip, width=\linewidth]{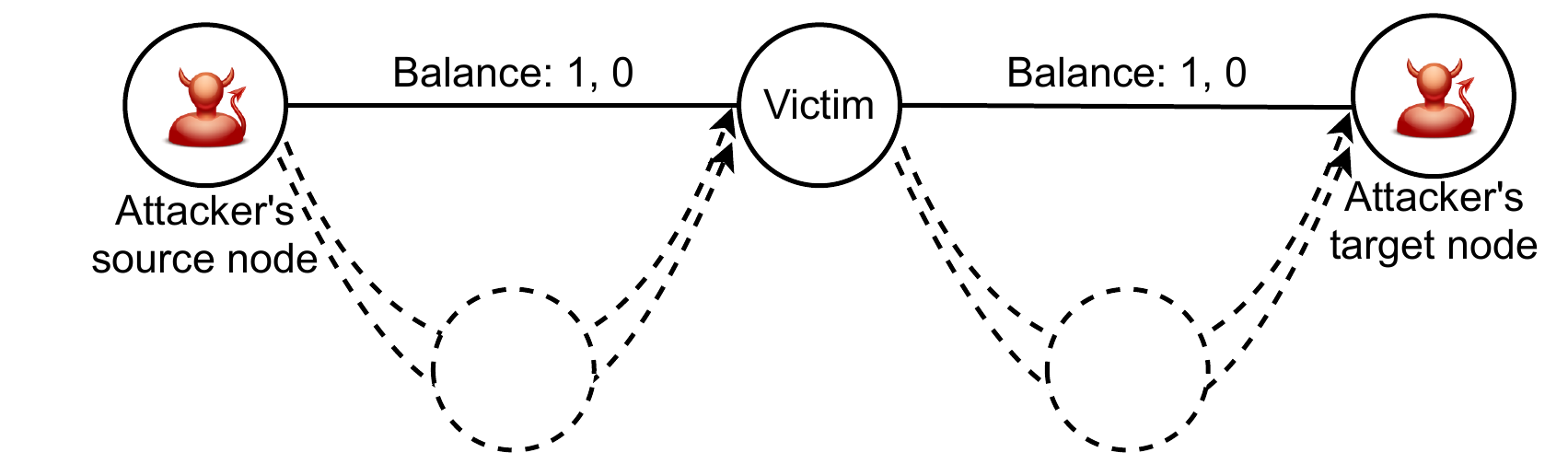}
		}
		\hfill
		\subfigure[Excessively loading channels with HTLCs]{
			\centering
			\includegraphics[trim = 290mm 297mm 130mm 80mm, clip, width=\linewidth]{./images/graphs/attack_floodLoot.pdf}
		}
		\hfill
		\subfigure[Resolving payments on the last hop]{
			\centering
			\includegraphics[trim = 290mm 232mm 130mm 140mm, clip, width=\linewidth]{./images/graphs/attack_floodLoot.pdf}
		}
		\hfill
		\subfigure[Waiting for all HTLCs to expire (approx. the same time) and collecting unspent outputs]{
			\centering
			\includegraphics[trim = 290mm 158mm 130mm 212mm, clip, width=\linewidth]{./images/graphs/attack_floodLoot.pdf}
		}
		\caption[]{Different stages of a flood and loot attack.}
		\label{figure:attack_floodLoot}
	\end{figure}~\\
	\par
	Importantly, a node can become a victim of flood and loot attack only if it opens a channel with the attacker's source node. \figurename{~\ref{figure:attack_floodLoot}} depicts \ankit{the} different steps of a simplified flood and loot attack. In the first step, the attacker establishes a number of channels from source node to target node via victim nodes. It is worth mentioning that a given victim can be participating in multiple distinct paths from source to target, which is much worse for such a victim node~(see \figurename{~\ref{figure:attack_floodLoot}}(a)). Now, source node commences (preferably when blockchain fees are low) multiple HTLC payments to the target node using the channels previously created~(see \figurename{~\ref{figure:attack_floodLoot}(b)}). The aim is to use as many channels as possible with maximum funds utilization. Next, the target node acts honestly, reveals the secret to the victim node, settles HTLC, and gets the payment~(see \figurename{~\ref{figure:attack_floodLoot}(c)}). The victim nodes \ankit{forward} the secret to the next node in the chain, which is source node in this case. Source node refuses to respond, leaving open HTLCs in the channels~(see \figurename{~\ref{figure:attack_floodLoot}(c)}). As the timeouts for HTLCs approach, the victim node tries to close the channels with the source node to claim all open HTLCs on the blockchain. Consequently, numerous transactions towards the parent blockchain are induced simultaneously~(see \figurename{~\ref{figure:attack_floodLoot}}(d)). Not all transactions will enter the blockchain due to congestion. At the same time, the replace-by-fee protocol will allow transactions offering higher fees to replace other conflicting transactions. Thus, the attacker raises the fee for his transactions to exceed those of the victims to claim the funds using replace-by-fee policy.\\
	\textit{Countermeasures:} There are two prominent countermeasures for this attack. The first one is to limit the maximum number of HTLCs that can remain unresolved at any point \ankit{in} time. This limitation will decrease the number of transactions competing to include in the blockchain in a given window of time. So, the attacker will require more victims to trigger the race conditions necessary for the attack. The second one is to \ankit{use reputation} scores, where channel opening requests from unknown counterparts is given a lower priority, which reduces the scope of locking larger funds in such channels. Another solution is to use anchor commitment output~\cite{AnchorOutputs} that manipulates the way transaction fees are paid. However, anchor outputs remain ineffective for an attacker with sufficiently large capital.

	\subsection{Griefing attack}
	\label{subsection:griefing_attack}
	Unlike most of the other attacks on \layerTwo protocols, griefing attack~\cite{griefing} is not directly motivated by stealing funds or information. It instead focuses on stalling payment networks by exhausting the channel capacity. It steals useful time of benign participants by preventing them from processing further transactions, which eventually leads to temporal loss of funds, decreased network throughput, and routing disruption.
	\begin{figure}[H]
		\centering
		\includegraphics[trim = 3mm 158mm 50mm 15mm, clip, width=\linewidth]{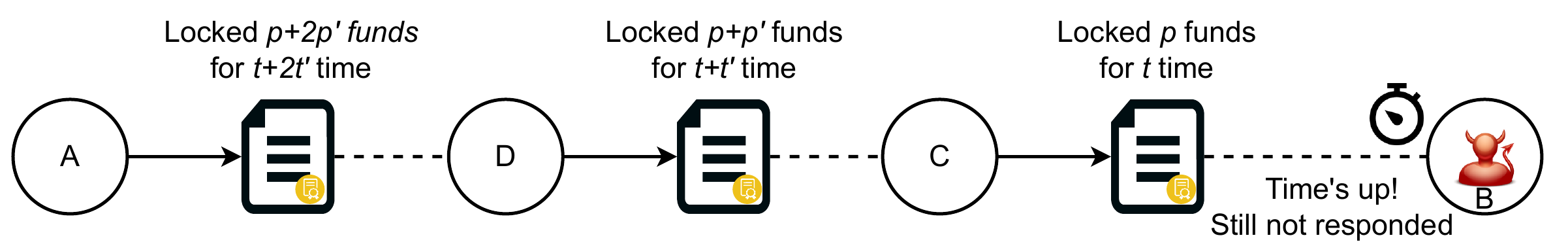}
		\caption[]{Griefing attack to stall network and participants.}
		\label{figure:attack_griefing}
	\end{figure}
	\figurename{~\ref{figure:attack_griefing}} illustrates a representative griefing attack setup, where A is transferring $p$~funds to B over a payment path A$\rightarrow$D$\rightarrow$C$\rightarrow$B, and each intermediate node charges a processing fee~$p'$. A forwards a conditional payment to D using an off-chain contract, where it locks~$p+2p'$ funds for a time period~$t+2t'$. Here, $t'$~indicates \ankit{the} deadline for confirmation time for on-chain settlement. D forwards the payment to C using another off-chain contract locking~$p+p'$ funds for a time period $t+t'$. C forwards the payment to B using another off-chain contract locking~$p$ funds for a time period~$t$. In order to claim~$p$ funds from C, B must resolve the payment within time period~$t$. Otherwise, C can claim a refund on-chain by closing its channel with B; similar procedures to be followed by D and then A. Nevertheless, C can go on-chain only after its contract with B expires. As a result, B can block $\mathcal{O}(p)$~funds in each of the preceding \ankit{contracts} for a time period~$t$ without losing any funds. It is worth mentioning that the order of~$t$ can be days for genuine users' convenience. Thus, each of the involved nodes can not utilize their funds for the entire duration of $t$. The attacker can stall the network by simultaneously launching multiple griefing attacks from its single or multiple nodes.\\
	\textit{Countermeasures:} Apart from limiting the number of incoming channels, faster contract resolution, and constant payment locktime, incentivizing/punishing nodes is a crucial way to tackle griefing attack. Griefing penalty~\cite{griefingPenalty} maneuvers along the same idea, which punishes the attacker to pay a penalty for compensating the victims' lost time. 
	This way it prevents nodes from ever attempting to participate in such an attack.

	\subsection{Time dilation/eclipse attack}
	\label{subsection:time_dilation_attack}
	Time dilation attack \cite{timeDilation} dilates a victim's clock. As a result, the victim always remains late in becoming aware of the new blocks in the chain. In other words, the victim misses updates from the network and keeps on working with outdated information. To this end, the attacker eclipses a victim node such that it cannot participate in the network owing to the delayed block delivery. Although time dilation attack is an expensive attack, it yields high returns in the form of stealing the total channel capacity in a single eclipse period. Authors in~\cite{routeHijacking, isolationDischarged} propose channel exhaustion and node isolation attacks along similar avenues. For the sake of brevity, we discuss only the time dilation attack.
	\begin{figure}[!htbp]
		\centering
		\includegraphics[trim = 3mm 3mm 160mm 102mm, clip, width=.8\linewidth]{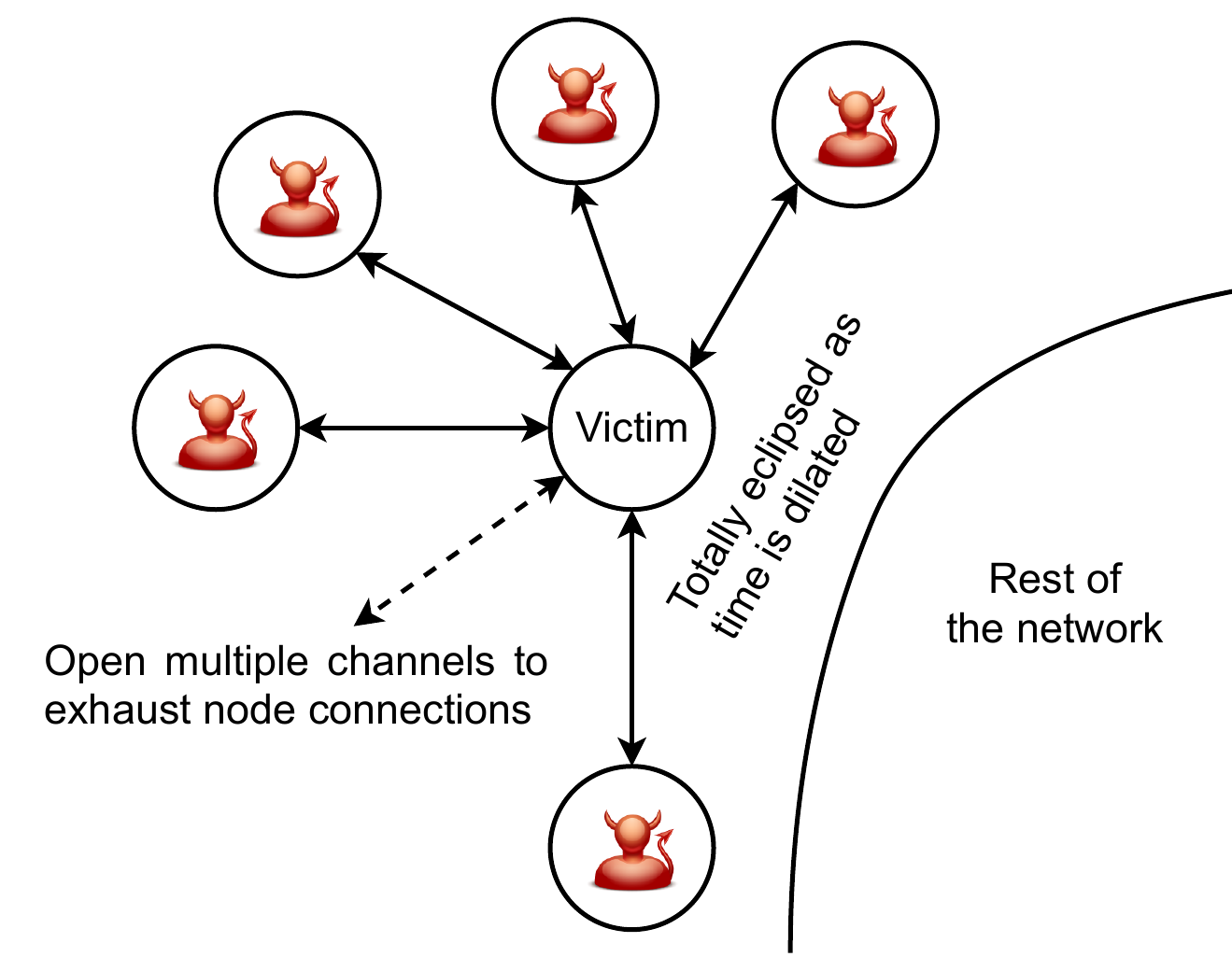}
		\caption[]{Eclipse attack where attacker dilates a user.}
		\label{figure:attack_eclipse}
	\end{figure}
	\par
	\figurename{~\ref{figure:attack_eclipse}} depicts multiple malicious nodes eclipsing a victim node. Typically, such an attack focuses on \ankit{a} single victim node. Essentially, eclipsing a user means preventing it from observing current network activities or state and cutting \ankit{its} communication with other honest users. 
	\ankit{To} do so, the attacker must occupy every connection a victim can have using pseudonymous nodes with pretend identities.	After eclipsing a victim, the attacker induces time dilation by introducing a delay between the time it receives a block to the time it will \ankit{feed} the block to the victim node. Time dilation attack is commonly seen in Lightning networks and can be classified - based on the target of the attack - into the following three main categories:
	\begin{itemize}
		\item \textit{Target channel's state finalization:} It pushes the victim's observed block height several blocks behind the tip of the main chain, typically by negotiating a new state committed to an outdated chain state.
		\item \textit{Target per-hop delay:} Once the attacker gets ahead of the victim's block height by dilation, it routes a payment through the victim, and at the same time finalizes the state of their channel onto the blockchain. It stops the victim from renegotiating through preimage disclosure, leaving no option and funds with the victim.
		\item \textit{Target packet finalization:} When a victim node knows the preimage for an incoming contract, but the remote peers do not respond on time, the victim node goes on-chain a few blocks before the expiration time to claim the contract. But, a dilated victim is bound \ankit{to} lose such a race despite responding ahead of time.
	\end{itemize}
	\textit{Countermeasures:} A solution to time dilation attack must detect as well as mitigate the attack. 
	Detecting whether a block was delayed in being sent or delivered is complex due \ankit{to high false positive rate}. Nevertheless, \ankit{an} abnormal routing failure rate could be an indicator for detection. Similarly, mitigating the attack is not straightforward mainly because identifying attacker committed state or choosing the right channels from multiple open channels is difficult. Watchtowers~\cite{watchtower1, watchtower2, watchtower3, watchtower4} act as a substitute \ankit{for a} built-in defense mechanism, but they come with \ankit{assumptions} on their honesty and efficiency.

	\subsection{Balance lockdown attack}
	\label{subsection:lockdown_attack}
	Balance lockdown attack~\cite{perez2020lockdown}, also known as balance availability attack, affects the ability of a victim node to successfully participate in payment routing. The attack impedes the victim node from taking part in any further transactions by blocking their balance funds. Such an attack confers a dominant position to the attacker, as well as reduces \ankit{the} system's efficiency. In particular, it enables the attacker to block certain \ankit{payment} paths, giving a competitive advantage to \ankit{attacker-favored} paths. The attacker has precise knowledge about network topology using which the attacker routes a payment via the victim node.
	\par
	A multihop payment in a channel network is atomic in nature, which means that the fund transfer can happen only after \ankit{the} establishment of the complete payment path. Thus, intermediate nodes on the payment path need to keep their funds locked. An attacker can lock an amount $p$ on each victim node on the payment path just by sending a payment of $p$ through them. It is worth noting that the cost of sending the payment and the time taken for the completion of the transaction are the two key factors that decide the feasibility of this attack. Hence, the attacker aims to increase the completion time for the transaction as well as to minimize the cost of the transaction. Authors in~\cite{perez2020lockdown} show that the overall effect of the attack can be captured by a parameter called Attack Effort Ratio (AER), which is defined as the ratio of capacity required to launch the attack and the capacity that the attack effectively blocks~(cf Eq.~{\ref{eq_AER}).
	\begin{equation}
		\label{eq_AER}
		AER = \frac{Capacity_{Required} }{Capacity_{Blocked}}
	\end{equation}
	Another parameter, called Total Blocked Time~(TBT), to measure attack's effectiveness is defined as the amount of time for which the funds of victims \ankit{are} blocked in the transaction.\\
	\textit{Countermeasures:} An optimal solution to handle balance lockdown attack should aim at increasing AER and reducing TBT. AER can be increased by disallowing loops in payment routes to minimize attacker's profitability. Another way to increase AER is to limit the maximum length of payment routes, though limiting route length may have \ankit{an} impact on the performance. Regulating TBT is tricky and has to traded-off with AER. Nevertheless, such trade-offs need detailed studies.
	
    \subsection{Balance discovery attack}
    \label{subsection:balance_disc}
    Balance discovery attack~\cite{herrera2019difficulty, dam2020improvements} affects PCN, in particular \ankit{the} Lightning network. A PCN \ankit{comprises} payment channels with a fixed deposit known as the channel capacity. The total capacity of a channel is publicly known. However, the individual balances on either side of a channel are not disclosed to preserve the privacy of users participating in the channel. Balance discovery attack aims to disclose individual balances of users, thus compromising users' privacy. Any node creating a channel with a malicious node is vulnerable to this attack. Moreover, the attack automatically extends to all the immediate neighboring nodes of that victim node.
    \begin{figure}[H]
    	\centering
    	\includegraphics[trim = 5mm 185mm 185mm 2mm, clip, width=.75\linewidth]{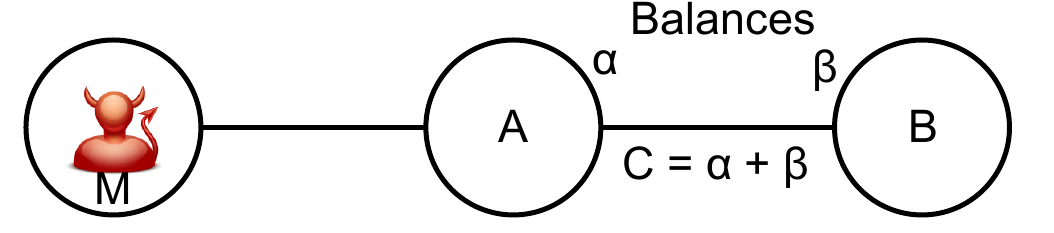}
    	\caption[]{Balance discovery attack where attacker tries to disclose the balance between
    		A and B.}
    	\label{figure:balance_discovery_attack}
    \end{figure}
    \par 
    \figurename{~\ref{figure:balance_discovery_attack}} illustrates the setup for balance discovery attack. Here, two nodes A and B form a channel with \ankit{a} total capacity $C$. We can safely assume that A and B perform some transactions via their channel over the period of time, and their current balances in the channel are $\alpha$ and $\beta$, respectively. It is important to mention that the total channel capacity remains unchanged, i.e., $C = \alpha + \beta$. A malicious node M wants to disclose $\alpha$ and $\beta$. To do so, M first creates a channel with A. This channel also facilitates M with a path to B through A. Now, M sends a payment $p$ to B through A. As long as $p$ is less than $\alpha$, it can be sent successfully through the route M$\rightarrow$A$\rightarrow$B. M increases the value of $p$ and sends a new payment to B. M repeats this process until an error occurs. 
    The last successful payment value $p$ reflects the approximate balance of A while $\beta$ can be computed by subtracting $\alpha$ from publicly known $C$. As an improvement, the value of $p$ can be efficiently calculated by using a binary search on lower and higher payment bounds. Furthermore, M can also send fake payments for probing such that none are finalized. Such \ankit{a} strategy helps in reducing \ankit{the} cost of the attack.\\
    \textit{Countermeasures:} One straightforward countermeasure is to limit the maximum allowed payment value. It would increase the efforts required by the attacker, but it would also affect the overall functionality of the system. The key approaches to tackle balance discovery attack are dropping rate parameter and dynamic absorption of negative balances~\cite{herrera2019difficulty}. The former suggests each node to randomly deny a specified percentage of transactions without disclosing the cause of failure to the payment sender. Such a strategy will deceive the attacker to assume that the payment has failed due to a lack of funds. Consequently, the attacker will interpret the wrong values of the balances. The latter advocates absorbing negative channel balances to prevent accurate probing of \ankit{the} channel's remaining capacities.
	
	\subsection{Congestion attack}
	\label{subsection:congestion_attack}
	Similar to balance lockdown attack, the goal of congestion attack~\cite{mizrahi2021congestion} is to block victims' funds at a lower cost. The attack targets PCNs and obstructs several payment channels at once for a long period of time. It exploits the trustless payment mechanism and long HTLC expiration duration. In particular, the attacker acts as both the source and destination of a payment. It can establish multiple such payments for wider coverage of the network. The attacker then sends funds along each of such routes, blocking nodes on each route.
	\begin{figure}[!htbp]
		\centering
		\includegraphics[trim = 0mm 5mm 53mm 2mm, clip, width=.9\linewidth]{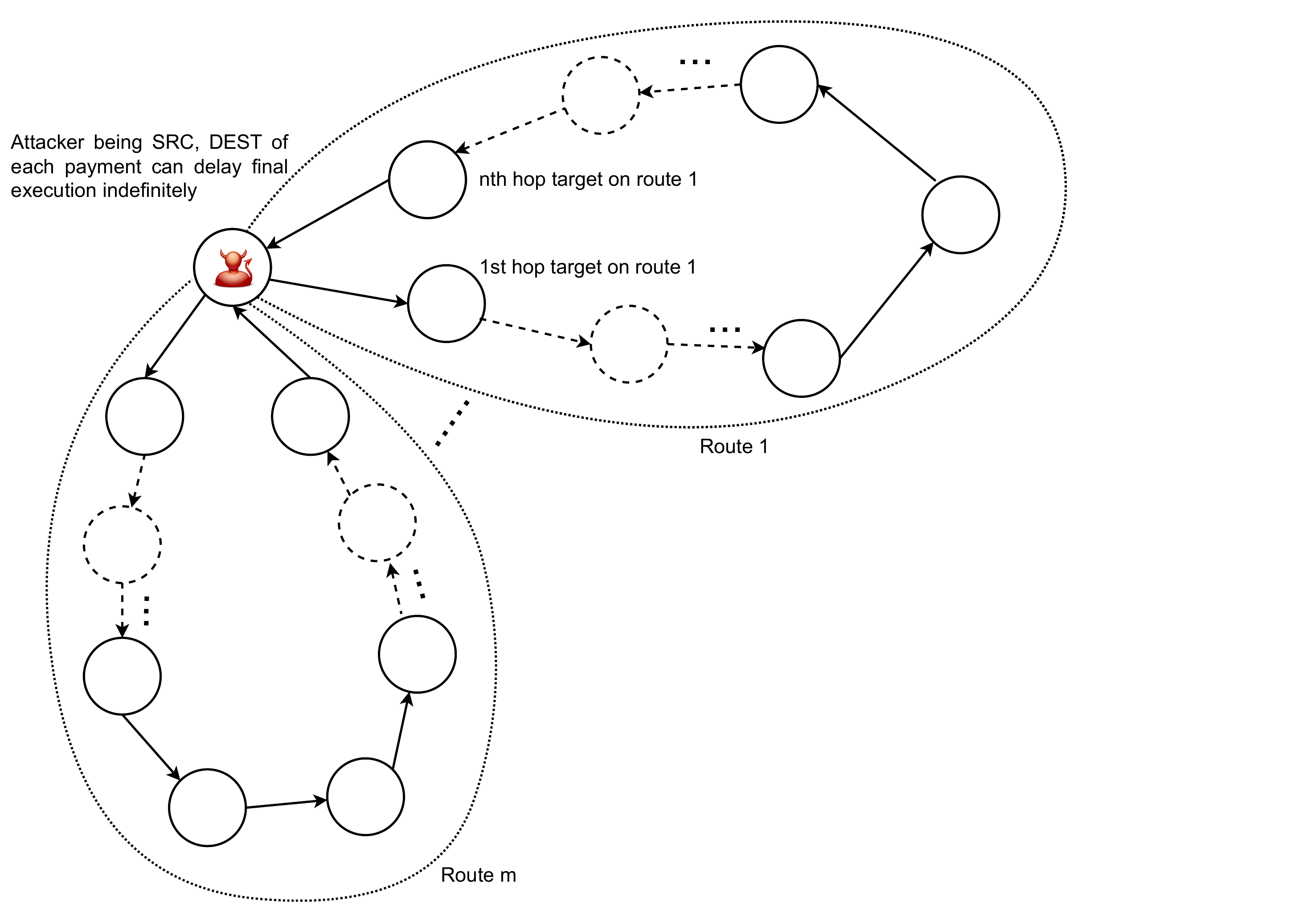}
		\caption[]{Congestion attack over $m$ payment routes.}
		\label{figure:attack_congestion}
	\end{figure}
	\begin{table*}[b]
		\centering
		\caption{A comparison of different \layerTwo solutions.}
		\label{table:comparison_discussion}
		\begin{tabular}{|c|l|c|c|c|c|}
			\hline
			\textbf{Criteria} & \multicolumn{1}{c|}{\textbf{Aspect}} & \textbf{\begin{tabular}[c]{@{}c@{}}Channels\end{tabular}} & \textbf{\begin{tabular}[c]{@{}c@{}}Commit\\ Chains\end{tabular}} & \textbf{\begin{tabular}[c]{@{}c@{}}Optimistic\\ Rollups\end{tabular}} & 
			\textbf{\begin{tabular}[c]{@{}c@{}}zk\\ Rollups\end{tabular}} \\ \hline
			\multirow{4}{*}{\textit{\begin{tabular}[c]{@{}c@{}}Performance\\ and\\ operations\end{tabular}}} & Transaction Speed & Fast & Moderate & Slow & Slowest \\ \cline{2-6} 
			& Transaction cost & Very low & Very low & Low & Low \\ \cline{2-6} 
			& \begin{tabular}[c]{@{}l@{}}On-chain transaction \\ for account opening\end{tabular} & Yes & No & No & No \\ \cline{2-6} 
			& Capital-efficient & No & Yes & Yes & Yes \\ \hline
			\multirow{3}{*}{\textit{\begin{tabular}[c]{@{}c@{}}Temporal\\ requirements\end{tabular}}} & Withdrawal time & \begin{tabular}[c]{@{}c@{}}One\\ confirmation\end{tabular} & \begin{tabular}[c]{@{}c@{}}Multiple\\ days\end{tabular} & \begin{tabular}[c]{@{}c@{}}Multiple\\ days\end{tabular} & \begin{tabular}[c]{@{}c@{}}\textless{}10\\ minutes\end{tabular} \\ \cline{2-6} 
			& Finality & Instant & \begin{tabular}[c]{@{}c@{}}One\\ confirmation\end{tabular} & \begin{tabular}[c]{@{}c@{}}One\\ confirmation\end{tabular} & \begin{tabular}[c]{@{}c@{}}\textless{}10\\ minutes\end{tabular} \\ \cline{2-6} 
			& \begin{tabular}[c]{@{}l@{}}Instant transaction confirmations\\ with full security guarantees\end{tabular} & \begin{tabular}[c]{@{}c@{}}Yes,\\ self supported\end{tabular} & \begin{tabular}[c]{@{}c@{}}No,\\ parent-linked\end{tabular} & \begin{tabular}[c]{@{}c@{}}No,\\ parent-linked\end{tabular} & \begin{tabular}[c]{@{}c@{}}No,\\ parent-linked\end{tabular} \\ \hline
			\multirow{4}{*}{\textit{\begin{tabular}[c]{@{}c@{}}Security\\ considerations\end{tabular}}} & Cryptographic primitives used & Standard & Standard & Standard & New \\ \cline{2-6} 
			& User liveness & Yes & Yes & Delegable & No \\ \cline{2-6} 
			& Mass exiting & No & Yes & No & No \\ \cline{2-6} 
			& Fund freezing by validators & No & No & No & No \\ \hline
			\multirow{3}{*}{\textit{\begin{tabular}[c]{@{}c@{}}Support,\\ options,\\ and\\ miscellaneous\end{tabular}}} & Smart contract support & Yes, limited & Yes, limited & Yes, flexible & Yes, flexible \\ \cline{2-6} 
			& \begin{tabular}[c]{@{}l@{}}Imports existing\\ EVM-bytecode\end{tabular} & No & No & Yes & Yes \\ \cline{2-6} 
			& Privacy options & Limited & No & No & Full \\ \hline
		\end{tabular}
	\end{table*}
	\par
	\figurename{~\ref{figure:attack_congestion}} depicts a congestion attack on $m$ payment routes, where all nodes along the route $1$ to $m$ are blocked until the attacker's transaction completes. After joining the network, the attacker evaluates different routes considering maximum route length, available funds, contract expiration limits, etc. Next, it establishes channels with two nodes along a target route such that one of the nodes sees the attacker as the source while the \ankit{other node} sees the attacker as the destination of the payment. The attacker can delay a transaction on such a route because it is both the source and destination of the payment, which leaves the nodes being blocked. Right before the HTLC expiration time, the attacker cancels the transaction and restores the channels to their original state. At this point, it can relaunch the attack to block the same path till the next HTLC expiry, which \ankit{can} be in the order of weeks. Importantly, this attack can be used in three ways: (1)~to block high liquidity channels, (2)~to isolate multiple node pairs, and (3)~to disconnect individual nodes from the network.\\	
	\textit{Countermeasures:} A simple approach to increase the number of HTLCs supported by benign nodes would remain ineffective as the attacker can take over all HTLCs with a larger volume of payments. Nevertheless, the effect of the attack can be limited by reducing the maximum allowed route length. Furthermore, the nodes can regulate the number of maximum concurrent payments with a peer based on its behavior.

	\section{Discussion}
	\label{section:discussion}
	Different \layerTwo solutions aim to scale blockchains while offering distinct functionalities based on different principles and underlying technologies. In this section, we compare the key categories of \layerTwo solutions. We have omitted cross chains and hybrid solutions as properties of cross chains implementations are highly dependent on chains targeted for interoperability while hybrid solutions have diverse hardware requirements. Thus, generalizing these categories may not truly reflect their characteristics. \ankit{We compare these solutions over multiple attributes that can be organized into four categories, which are performance, temporal requirements, security considerations, and miscellaneous options. \tablename{~\ref{table:comparison_discussion}} presents a summary of the comparisons.}
	\par
	In terms of performance, channels offer faster transaction processing (i.e., almost instant \layerTwo transfers) at a meager cost of transactions. However, channels require locking collaterals on-chain, while other solutions may not enforce this requirement. In particular, each channel must have collateral locked to start the operations. Thus, other solutions are perceived as more capital efficient than channels.
	\par
	Withdrawing balances from channels requires just one on-chain confirmation. Commit chains and Optimistic Rollups typically elongate withdrawal time for balance security and dispute resolutions. Such longer durations can be shortened by introducing risk insurers or liquidity providers. However, the reliability of such insurers has its own concerns. Withdrawing the balance on \layerOne from zk~Rollups takes fewer minutes in the best case. The finality of a transaction reflects a state, where the transaction can not be reverted on the main chain. The underlying principles of channels offer instant finality while other solutions typically offer delayed finality. Nevertheless, instant finality can be tweaked on other \layerTwo solutions as well to reflect instant confirmation to \ankit{the} user, but full security guarantees remain exclusive to channels.
	\par
	Most of \ankit{the} \layerTwo protocols use standard cryptographic primitives, SNARKs and STARKs are heavily used in \ankit{zero-knowledge-based protocols}. User liveness assumption is critical, which can be defined as the requirement of a user to remain online to receive/verify transactions, monitor disputes, and handle misbehaving counterparties. Unlike channels, commit chains enable users to receive \ankit{transactions} while remaining offline. However, users are advised to come periodically online to inspect checkpoint commitment. This assumption about a user's online status can be delegated to a trusted third-party that monitors transactions on behalf of the user. Nonetheless, such a third-party may compromise if the incentive to misbehave is greater than its guarantee deposit. Only commit chains allow all users to successfully withdraw - mainly for security reasons - in a short period of time, and all the protocols prevent validators from confiscating \ankit{the} funds of participants.
	\par
	Both Optimistic and zk~Rollups offer support importing existing EVM-bytecode with minor modifications, and thus, have a flexible support for smart contracts. Transaction deanonymization and user profiling are major privacy concerns that only zk~Rollups address by default~\cite{deanon1, deanon2, matterlabsComparison}.
	
	\section{Conclusion}
	\label{section:conclusion}
	Scalability is a major issue for blockchain-based solutions. Vast efforts, from both academia and industry, have been put in different directions to tackle the blockchain scalability issue. Such efforts have resulted in a rich literature of \layerTwo protocols that primarily aim to scale underlying main chains. In this work, we first create a broader taxonomy of \layerTwo protocols, which is followed by a detailed explanation of each \layerTwo protocol class. These protocols bring scalability to the main chains at the cost of different security assumptions and guarantees. Thus, we discuss various issues associated with these protocols and also compare them against each other. We believe that our study offers better explanations and analyses to help the readers understand the domain better.


\bibliographystyle{IEEEtran}
\balance
\bibliography{bib}
\balance

\appendices
\counterwithin{table}{section}
\renewcommand{\thesection}{\Alph{section}}%


\end{document}